\documentclass[aps, reprint, groupeaddress]{revtex4-1}
\usepackage[utf8]{inputenc}
\usepackage[english]{babel}
\usepackage{physics}
\usepackage{float}
\usepackage[normalem]{ulem}

\usepackage{blindtext}
\usepackage{times}
\usepackage{graphicx,epsfig}
\graphicspath{{images/}} 
\usepackage{color}
\usepackage[usenames,dvipsnames]{xcolor}
\usepackage{amsmath,bbm,amssymb, amsthm}
\usepackage{dsfont} 
\usepackage{stmaryrd}
\definecolor{linkcolor}{rgb}{0,0,0.6}		
\definecolor{bleu}{HTML}{1732a6}
\usepackage[colorlinks=true, pdfstartview=FitV, linkcolor= linkcolor, citecolor= linkcolor, urlcolor= linkcolor, hyperindex=true, hyperfigures=false]{hyperref}
\usepackage{numprint}

\addto\captionsenglish{}

\newcommand{\la}{\langle}
\newcommand{\ra}{\rangle}
\newcommand{\mc}{\mathcal}

\newcommand{\taug}{\mathcal{T}}

\begin{document}

\title{Gambling Carnot Engine}

\author{Tarek Tohme$^{1,2\star}$, Valentina Bedoya$^{1\star}$,  Costantino di Bello$^3$,  L\'ea Bresque$^{1}$, Gonzalo Manzano$^4$, and \'Edgar Rold\'an$^{1,*}$, \medskip}

\address{$^1$ICTP -- The Abdus Salam International Centre for Theoretical Physics, Strada Costiera 11, 34151 Trieste, Italy\\
$^2$Laboratoire de Physique de l’\'Ecole Normale Sup\'erieure, CNRS,
PSL University, Sorbonne Universit\'e, and Université de Paris, 75005 Paris, France \\
$^3$University of Potsdam, Institute of Physics \& Astronomy, 14476 Potsdam, Germany\\
$^4$Institute for Cross-Disciplinary Physics and Complex Systems, (IFISC, UIB-CSIC),
Campus Universitat de les Illes Balears E-07122, Palma de Mallorca, Spain}

\begin{abstract}
We propose a theoretical model  for a colloidal heat engine driven by a feedback protocol that is able to fully convert the net heat absorbed by the hot bath into extracted work. 
The feedback protocol, inspired by gambling strategies, executes a sudden quench at zero work cost when the particle position satisfies a specific first-passage condition.
As a result, the engine enhances both power and efficiency with respect to a standard Carnot cycle, surpassing Carnot's efficiency at maximum power. Using first-passage and martingale theory,  we derive analytical expressions for the power and efficiency far beyond the quasistatic limit and provide scaling arguments for their dependency with the cycle duration. Numerical simulations are in perfect agreement with our theoretical findings, and illustrate the impact of the data acquisition rate on the engine's performance.    
\end{abstract}

\maketitle

\textit{Introduction}. A central goal for statistical physics in the XXI century, through the development of {\em stochastic thermodynamics}~\cite{sekimoto1998langevin,peliti2021stochastic}, is the search of efficient protocols of energy harvesting from the fluctuations of small systems. 
So far, the combination of theory and experiments in stochastic thermodynamics have led to two key research programs along efficient energy extraction: (i) mimicking classical engines with miniaturized analogues (e.g. stochastic heat engines~\cite{schmiedl2007efficiency}), and (ii) exploiting information processing through feedback control protocols (e.g. Maxwell's demons~\cite{LeffRex}).  

Two hundred years after Carnot's {\em Reflections sur la puissance motrice du feu}~\cite{carnot1878reflexions}, the realization and characterization of nanometer-sized analogues of e.g. Carnot's~\cite{martinez2016brownian}, Stirling's~\cite{blickle2012realization}, and Otto's~\cite{rossnagel2016single} cycle  have become a  frontier topic of active research~\cite{krishnamurthy2016micrometre,martinez2017colloidal,martin2018extracting}. In particular, it has been shown that the second law of stochastic thermodynamics applied to such {\em stochastic heat engines}~\cite{schmiedl2007efficiency} implies the  Carnot bound on the average efficiency of any of such engines. More precisely,
\begin{equation}
    \eta\equiv -\frac{\langle W\rangle}{\langle Q_{\rm h}\rangle}\leq 1-\frac{T_{\rm c}}{T_{\rm h}},\label{eq:Sadi_Carnot}
\end{equation}
i.e. the efficiency cannot exceed Carnot's efficiency~$\eta_C=1-(T_{\rm c}/T_{\rm h})$, with $T_{\rm c}$ and $T_{\rm h}>T_{\rm c}$ the temperature of the cold and the hot bath, respectively. 
In Eq.~\eqref{eq:Sadi_Carnot}, $-\langle W\rangle$ and $\langle Q_{\rm h}\rangle$ denote the work extracted from the engine over a cycle and the heat absorbed by the engine from the hot bath, respectively, both averaged over many realizations of the same deterministic cyclic protocol~\cite{note}.

In parallel, fruitful theoretical and experimental advances have investigated thermodynamic cycles driven by feedback-control protocols, as inspired from the celebrated Maxwell's demon~\cite{parrondo2015thermodynamics,Lutz15,Koski15,Camati16,Paneru18,Paneru2020,Saha21,Fadler23,Saha23}. Here, an external agent (e.g. a ``demon") retrieves information from the stochastic evolution of a system. Such information  may be processed by the demon in  clever ways to control the system via feedback strategies. Generalizations of the second law of stochastic thermodynamics~\cite{sagawa2010generalized} have revealed that the Carnot bound~\eqref{eq:Sadi_Carnot} does not hold necessarily in the presence of feedback control. When feedback control is applied to heat engines, second laws of thermodynamics with feedback control apply~\cite{Sagawa09,sagawa2010generalized,horowitz2010nonequilibrium,Toyabe2010,Sagawa12,Horowitz14,barato2014unifying,datta2022second}, which imply that
$  -\langle W\rangle /(\langle Q_{\rm h}\rangle +k_{\rm B}T_{\rm c} \langle I\rangle/\eta_C)\leq \eta_C$, where  $ k_{\rm B}$ is Boltzmann's constant, and $\langle I\rangle$ is an information-theoretic quantity that can be positive or negative depending on whether information flows from the demon to the engine, or vice versa. So far, it remains unclear whether there exist feedback protocols that, when applied to realistic heat engines, enable to fully convert the heat absorbed from the hot bath into extracted work. In other words, how can one design an optimal feasible feedback protocol of full heat-to-work conversion at the expense of information, for which $-\langle W\rangle=\langle Q_{\rm h}\rangle$, and thus the efficiency parameter defined in Eq.~\eqref{eq:Sadi_Carnot} gives $\eta=1$? 

In this Letter, we introduce an experimentally realizable feedback protocol (denoted the ``Gambling Carnot Engine", GCE), that is able to fully convert cyclically the net heat absorbed from a hot bath into work extracted on the cycle. In other words, such protocol obeys
\begin{equation}
    \eta_{\rm G}\equiv -\frac{\langle W\rangle_{\rm G}}{\langle Q_{\rm h}\rangle_{\rm G}}\leq 1,
    \label{eq:SadiGambler}
\end{equation}
where here and further the subindex G denotes averages over many realizations of the GCE. 
Remarkably, we find that the inequality~\eqref{eq:SadiGambler} 
saturates in the limit of large cycle times. The GCE imbibes wisdom from the recently introduced gambling demons~\cite{manzano2021thermodynamics} and continuous Maxwell demons~\cite{rico2021dissipation}, where an external controller executes a 
feedback operation
as soon as a prescribed condition is first met. This class of demons require continuous monitoring schemes in order to apply the feedback strategy successfully, and so far they have mostly been studied in isothermal scenarios, both theoretically and experimentally~\cite{manzano2021thermodynamics,rico2021dissipation,dago2022virtual,du2024performance}. The GCE that we develop in this work consists of the combination of the Brownian Carnot engine (BCE)~\cite{martinez2016brownian}) with a gambling 
demon that exerts a sudden shortcut in the isothermal compression 
when the engine reaches a vanishing generalized pressure, and that shortcut is thus done at zero work expenditure. As we show below, such a protocol can outperform the BCE in both power and efficiency, and go beyond the Carnot limit.

{\em Gambling protocol}. We introduce a stochastic model for a Gambling Carnot Engine (GCE) where an external observer uses the information  acquired on the working substance ---a Brownian particle--- up to a prescribed deadline time, to perform a feedback protocol that results in an enhancement 
 of the average work extracted per cycle with respect to the BCE. 
The GCE  operates cyclically between two different temperatures, one cold $T_{\rm c}$ and another hot $T_{\rm h}>T_{\rm c}$. 
Each cycle  consists of four steps: (I) isothermal compression, (II) adiabatic compression, (III) isothermal expansion and (IV) adiabatic expansion. 
It is during the first step, (I), operating at 
$T_{\rm c}$, that gambling may occur, i.e., that a feedback  operation
may be applied on a cycle based on  information acquired from the working substance. 
Experimentally, the GCE protocol can be applied to 
an underdamped Brownian particle of mass $m$ that is immersed in a 
viscous fluid with friction coefficient $\gamma$. 
The particle is  trapped by a time-dependent harmonic potential $U_t(x) = \kappa_t x^2/2$, where~$\kappa_t$ is the trap stiffness at time $t$.
It is by changing this stiffness that the expansion and compression are implemented.
Moreover,   one can effectively vary the time-periodic temperature $T_t$ by e.g. changing the amplitude of an external white noise applied to the particle~\cite{martinez2015adiabatic}.

\begin{figure}[t]
    \centering
\includegraphics[width=0.85\columnwidth]{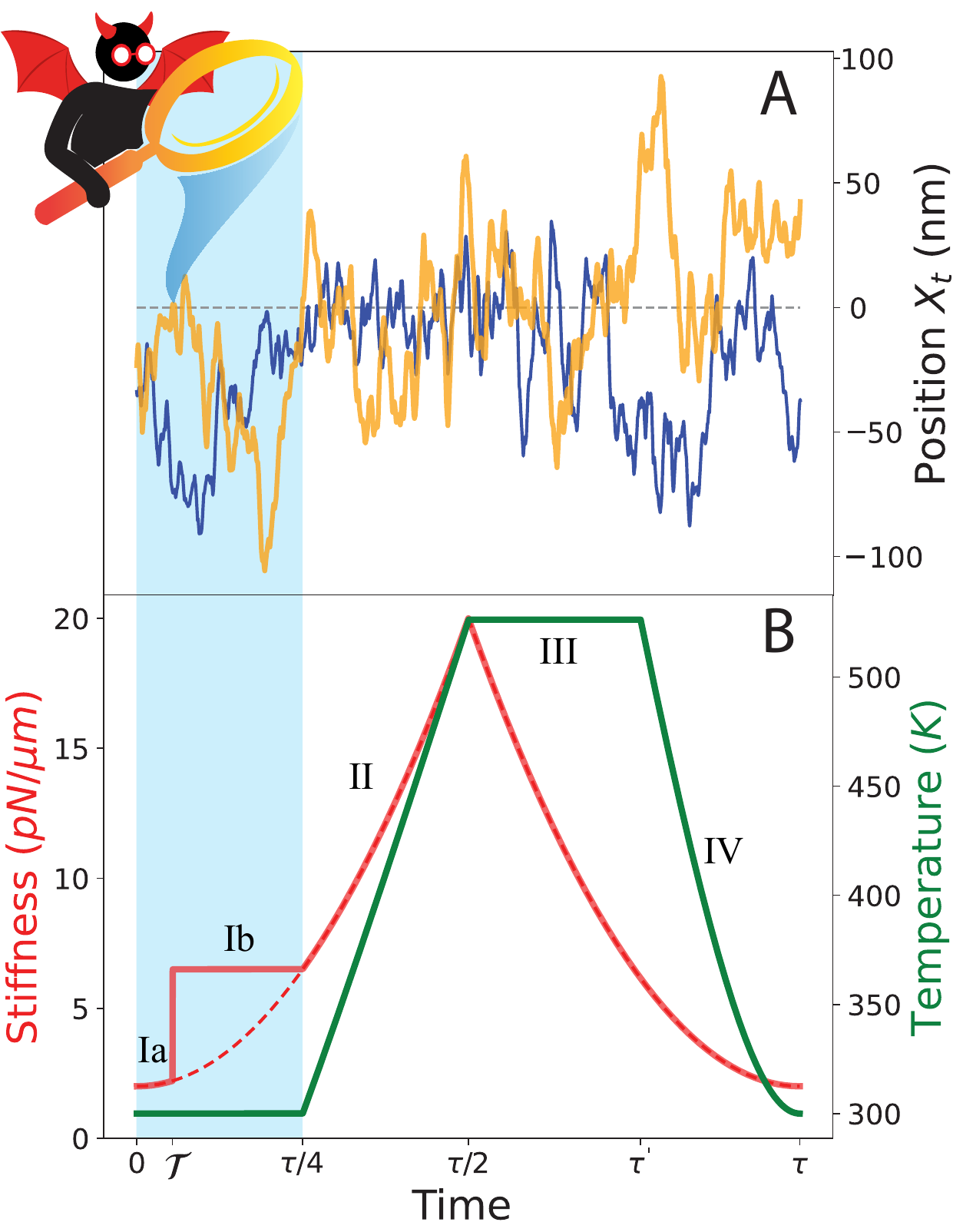}
    \caption{(A) Sketch of a gambling Carnot engine (GCE) at work together with its driving
    protocol (B) along a cycle of duration $\tau$.  A ``demon"  monitors the fluctuating position of an underdamped, trapped Brownian particle (panel A) 
    during the isothermal compression (blue light square) 
     following the stochastic feedback protocol in eqs.~\eqref{eq:kappaC},\eqref{eq:TC} (panel B).
     If the particle first crosses the origin  at a stochastic time $\mathcal{T}<\tau/4$ 
     (orange trajectory in panel A), the stiffness is switched suddenly at time $\mathcal{T}$ to 
     $\kappa_{\tau/4}$
     (continuous red line in panel B). On the other hand, if the particle does not cross the origin before $\tau/4$ (blue trajectory in panel A), the stiffness follows the deterministic protocol from the experimentally-realized Brownian Carnot engine~\cite{martinez2016brownian} (dashed red line in panel B). 
     } 
    \label{fig1}
\end{figure}

During a cycle, the particle position $X_t$ in the GCE follows a one-dimensional underdamped Langevin equation
\begin{equation}
    m\ddot{X}_t = - \gamma \dot{X}_t - \kappa_t X_t + \sqrt{2\gamma k_{\rm B}T_t}\,\xi_t,
    \label{eq:lang}
\end{equation}
where the dot 
denotes derivative with respect to time, and $\xi_t$ is a Gaussian white noise with zero mean $\langle\xi_t\rangle=0$ and autocorrelation $\langle\xi_t\xi_{t'}\rangle=\delta(t-t')$. The trap stiffness follows the stochastic  driving protocol 

\begin{equation}
    \kappa_t = \begin{cases}
        \kappa_0 + \alpha t^2&  0\leq t < \taug \qquad\; \text{(Ia) Monitoring} \\
        \kappa_0 + \alpha (\tau/4)^2 & \taug\leq t < \tau/4 \quad\; \text{(Ib) Waiting}\\
        \kappa_0 + \alpha t^2 & \tau/4\leq t < \tau/2 \quad \text{(II)}\\
        \kappa_0 + \alpha (\tau - t)^2  & \tau/2\leq t \leq  \tau  \quad \text{(III) and (IV)}
    \end{cases} 
\label{eq:kappaC}
\end{equation}
where the feedback operation is included in the isothermal compression (I) by introducing a sudden quench at the stochastic ``gambling time" $\taug$. Here, $\tau$ is the  total cycle duration time, $\kappa_0$ the initial  stiffness, 
and  $\alpha=4(\kappa_{\tau/2}-\kappa_{0})/\tau^2$. Note that $\kappa_0$, $\kappa_{\tau/2}$ 
and $\tau$ (and thus also $\alpha$) are fixed deterministic parameters.  On the other hand, $\taug$ 
is a random variable that takes on a different value in each cycle. It is formally defined as
\begin{equation}
\label{stop_time}
    \taug = \min (\mathcal{T}_0,\tau/4),
\end{equation}
 with $\mathcal{T}_0 = \inf \{t\geq 0:X_t=0\}$ 
the first-passage time to reach the origin 
 of positions (the trap centre). In other words, $\taug$  equals the first time at which the particle position changes sign if it occurs before the deadline $\tau/4$, and $\tau/4$ otherwise.
Figure~\ref{fig1} illustrates how the GCE works for two example trajectories in which one crosses the origin before the deadline (orange line) while the other does not (blue line). 

For the former, a sudden quench of the stiffness is performed at $\taug$ (red dashed line in Fig.~\ref{fig1}B), while for the latter the stiffness varies exactly as in the BCE (red solid line in Fig.~\ref{fig1}B). 

The adiabatic steps II and IV are implemented by keeping the ratio $T_t^2/\kappa_t$ constant, which 
enforces system entropy conservation in the quasistatic limit~\cite{martinez2016brownian}. 
Consequently, the temperature varies in time in the GCE, as in the BCE, according to (see green line in Fig.~\ref{fig1}B): 
\begin{align}
    T_t = \begin{cases}
        T_{\rm c}& 0 \leq t < \tau/4\qquad \text{(I)} \\
        T_{\rm c}\sqrt{\displaystyle\frac{\kappa_0 + \alpha t^2}{\kappa_{\tau/4}}}  & \tau/4 \leq t < \tau/2 \quad \text{(II)} \\
        T_{\rm h}&\tau/2 \leq t <\tau'\;\;\quad \text{(III)} \\
        T_{\rm c}\sqrt{\displaystyle\frac{\kappa_0 + \alpha (\tau-t)^2}{\kappa_0}}& \tau' \leq t <\tau. \;\;\qquad \text{(IV)} 
    \end{cases}
    \label{eq:TC} 
\end{align}
As shown in Refs.~\cite{martinez2015adiabatic,martinez2016brownian,barato2018arcsine}, the temperature of the hot bath $T_{\rm h}$ and the time $\tau'$  are constrained to specific values 
in order to ensure 
a vanishing heat transfer on average during the adiabatic compression (II) and adiabatic expansion (IV). In other words, $T_{\rm h}$ and $\tau'$ are functions of $T_{\rm c}$, $\tau$, $\kappa_0$, $\kappa_{\tau/2}$, and $\kappa_{\tau/4}$.   Such constraints imply that 
$T_{\rm h} = T_{\rm c} \sqrt{\kappa_{\tau/2}/\kappa_{\tau/4}}$ and $\kappa_{\tau'} = \kappa_0 + \alpha (\tau- \tau')^2  =( \kappa_0 \kappa_{\tau/2})/\kappa_{\tau/4}$, which using Eq.~\eqref{eq:kappaC} 
 gives $\tau'\simeq 0.76\tau$.



\begin{figure}[t] 
    \centering
    \includegraphics[width=0.85\columnwidth]{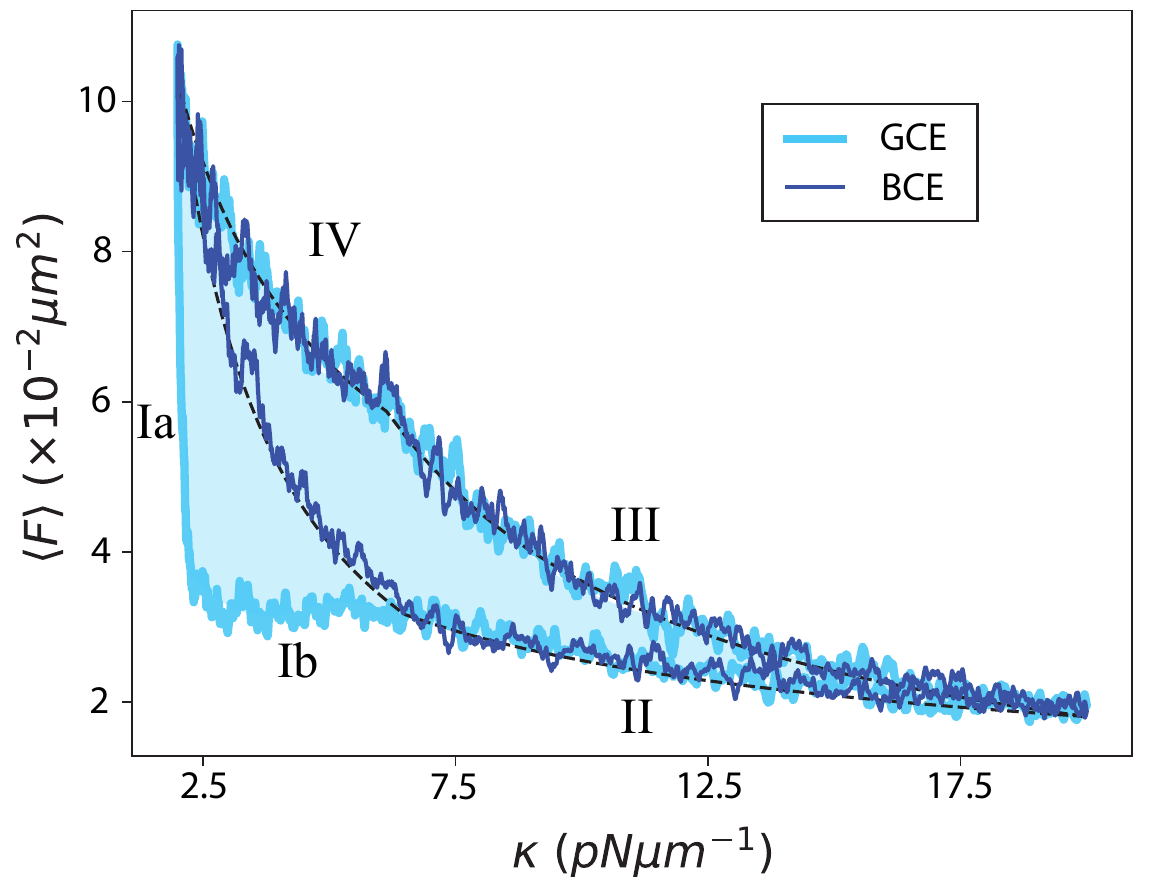}
    \caption{ 
    Clapeyron diagram $\langle F \rangle-\kappa$ displaying the  generalized force $F$ as a function of  the control parameter $\kappa$ ---with F averaged over 1000 cycle realizations of duration $\tau$ = 200 ms--- for the GCE (cyan thick line) and BCE (blue thin line). 
    The black dashed line is the Clapeyron diagram associated with the   BCE in the quasistatic limit. The area within shaded in light blue is the average work extracted by the  GCE. 
    Simulation parameters mimicked realistic experimental scenarios done with optically trapped polystyrene spheres in water~\cite{martinez2017colloidal}.
}
    \label{fig2}
\end{figure}

\textit{Stochastic thermodynamics.} 
The stochastic work exerted on the particle in $[t,t+dt]$ reads $dW_t=(\partial_t U_t)dt= (1/2)X_t^2 \dot{\kappa}_t dt$. Equivalently, we can write $dW_t=F_t d\kappa_t$, with $F_t= (1/2)X_t^2 $ the generalized force exerted on the particle and $d\kappa_t=\kappa_{t+dt}-\kappa_t$ the generalized displacement in $[t,t+dt]$. At the gambling time, $X_\taug=0$ and thus $dW_\taug\simeq 0$. Therefore, the stochastic work done on the particle along step~I
reads 
$ W_{\rm I} = W_{\rm Ia} + W_{\rm Ib}$. Neglecting the work exerted in the quench and noting that 
$ W_{\rm Ib} =0$ because $d\kappa_t=0$ after the quench, we have that the stochastic work done on the particle in step I reads
$ 
     W_{\rm I}\simeq \frac{1}{2}\int_0^{\taug}     X^2_t  d\kappa_t$ .     
Notably, because $X^2_t\geq 0$ and $d\kappa_t\geq 0$ along the isothermal compression, we have $ W_{\rm I}\leq \frac{1}{2}\int_0^{\tau/4}     X^2_t  d\kappa_t $, thus  gambling  leads 
to smaller work expenditure in step (I) than if no gambling were applied.
Such saving in work expenditure is larger the earlier the quench occurs and vanishes only for trajectories for which $\taug=\tau/4$. 
Figure~\ref{fig2} shows  a comparison between the Clapeyron diagrams of the GCE and BCE, which reveal a larger area enclosed on average by the GCE trajectories in the $\langle F\rangle $-$\kappa$ diagram. Thus, a key consequence of gambling is the reduction of the average work exerted in a cycle, $\langle W\rangle_{\rm G}\leq \langle W\rangle$, i.e. a larger work extraction per cycle in the GCE than in the BCE. 



\begin{figure}[ht]
    \centering
\includegraphics[width=0.85\columnwidth]{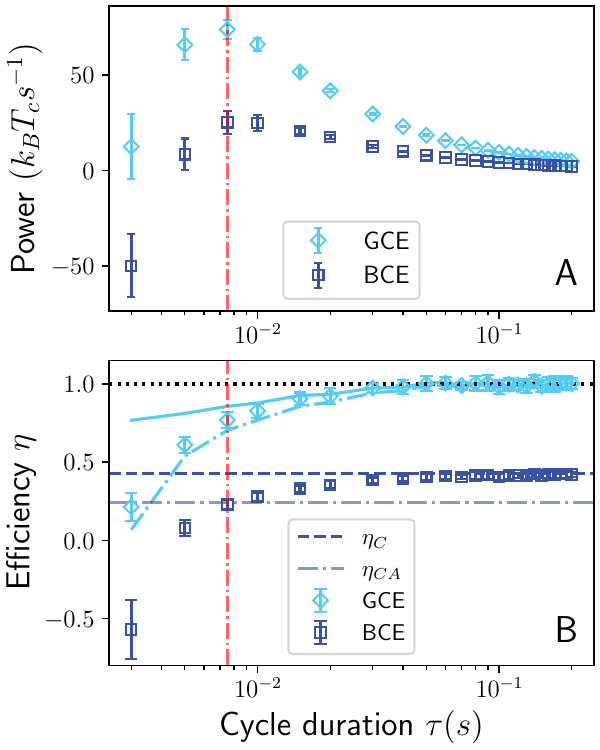}
    \caption{
    Power (A) and efficiency (B) of the GCE (light blue diamonds) and the BCE (dark blue squares)  as a function of the cycle duration~$\tau$, obtained from  numerical simulations of Eq.~(\ref{eq:lang}) and (\ref{stop_time}). 
    Cyan lines in (B) are numerical estimates for the upper (solid line) and lower (dash dotted line) bounds for the efficiency given by  Eq.~\eqref{eq:etabounds}. The horizontal lines are reference efficiency values: Carnot  $\eta_C=1-(T_{\rm c}/T_{\rm h})$ (dashed line),  Novikov-Curzon-Ahlborn  $\eta_{CA}=1-\sqrt{T_{\rm c}/T_{\rm h}}$ (dot-dashed line),  and  efficiency one (dotted line).
    For each cycle time, error bars are one-standard-deviation confidence intervals obtained from 10 simulations of $3\times 10^3$  cycles each, with fixed simulation time step  $\Delta t=0.2\, \mu$s, and  100KHz sampling frequency to approximate experimental conditions. The simulation parameters are: minimum stiffness $\kappa_0 = 2$ pN/$\mu$m, maximum stiffness $\kappa_{\tau/2} = 20$ pN/$\mu$m, cold temperature $T_{\rm c} = 300$ K, hot temperature $T_{\rm h} = 525$ K, mass $m = 0.5$ fg, and friction coefficient $\gamma = 2$nN$\mu$s/$\mu$m. In our simulations we assume instantaneous gambling such that we neglect the work done during a quench, and thus omit it in the calculation of efficiency.}
    \label{fig:3}
\end{figure}

{\em Power and efficiency}. To quantify the impact of gambling in the GCE's performance, we  evaluate in numerical simulations $\eta_{\rm G} = -\langle W\rangle_{\rm G}/\langle Q_{\rm h}\rangle_{\rm G}$ as a quantifier for the efficiency   [see Eq.~\eqref{eq:SadiGambler}] and $P_{\rm G} = -\langle W\rangle_{\rm G}/\tau $ for the power of a GCE with cycle time $\tau$ ranging from~$3\,$ms to $200\,$ms, see Fig.~\ref{fig:3}. 
As a benchmark, we compare $\eta_{\rm G}$ and $P_{\rm G}$ with the efficiency $\eta = -\langle W\rangle/\langle Q_{\rm h}\rangle$ and power $P= -\langle W\rangle/\tau $ obtained from numerical simulations of the BCE. Our numerical simulations reproduce the experimental results from 
Ref.~\cite{martinez2016brownian} for the BCE: (i) the efficiency $\eta\leq \eta_C$ and saturates to the Carnot efficiency $\eta_C$ for large cycle times $\tau>50$ms (dark blue squares in Fig.~\ref{fig:3}a); and (ii) the extracted power $P$ reaches its maximum value for cycle times $\tau$ at which the efficiency reaches the Novikov-Curzon-Ahlborn value $\eta_{\rm NCA}=1-\sqrt{T_{\rm c}/T_{\rm h}}$ 
(dark blue squares in Fig.~\ref{fig:3}b). Notably,  the GCE beats both the Carnot limit for the efficiency and the Novikov-Curzon-Ahlborn limit for the efficiency at maximum power. First, the efficiency $\eta_{\rm G}$ surpasses the Carnot limit for $\tau$ small (i.e. in fast cycles far from
equilibrium), see cyan diamonds in Fig.~\ref{fig:3}a. Even beyond, $\eta_{\rm G}$ approaches one from below while increasing $\tau$ towards the quasistatic limit, in which a $100\%$ conversion of the heat transfer from the hot bath into work extraction is achieved at the expense of gambling. 
Second, the average power extracted in the GCE verifies $P_{\rm G}\geq P$ for all values of $\tau$, as expected (cyan diamonds in Fig.~\ref{fig:3}b),  and the GCE's efficiency at maximum power  exceeds not only $\eta_{\rm NCA}$ but also the Carnot limit~$\eta_C$.

To gain a deeper understanding of GCE's  power and efficiency upgrades  with respect to the BCE, we first invoke recent results from the martingale thermodynamics theory~\cite{roldan2024martingales} valid for generic stopping times~\cite{manzano2021thermodynamics}. For moderate cycle time values $\tau$ yet larger than the characteristic relaxation times of the BCE, a reasonable assumption is that the position distribution in steps II, III and IV is unaffected by the gamble. Therefore, the average work in a cycle of the GCE exceeds that of the BCE by their difference in the isothermal compression, i.e. $\langle W\rangle-\langle W\rangle_{\rm G} \simeq \langle W_{\rm I} \rangle- \langle W_{\rm I} \rangle_{\rm G}$. Similarly we can also assume $\langle Q_{\rm h}\rangle_{\rm G}\simeq \langle Q_{\rm h}\rangle$ for moderate or large values of $\tau$. Under these assumptions, the efficiency of GCE is enhanced with respect to the efficiency of the BCE by 
\begin{equation}
    \Delta \eta \equiv \eta_{\rm G}-\eta 
    \simeq  \frac{\langle W_{\rm I} \rangle- \langle W_{\rm I} \rangle_{\rm G} }{\langle Q_{\rm h} \rangle}\geq 0.
\end{equation}
Applying the second law of thermodynamics at stopping times and related inequalities from martingale theory~\cite{manzano2021thermodynamics,roldan2024martingales,manzano2024thermodynamics} 
to the GCE,  
we derive the following lower and upper bounds for the efficiency enhancement: 
\begin{equation} \label{eq:etabounds}
\begin{split}
   \Delta \eta \geq \frac{\langle \Delta F_{\rm I} \rangle - \langle \Delta F_{\rm I}  \rangle_{\rm G}}{\langle Q_\mathrm{h} \rangle} + \frac{k_{\rm B} T_\mathrm{c} \langle \delta \rangle_{\rm G}}{\langle Q_\mathrm{h} \rangle}, \\
\Delta \eta \leq 
      \frac{\langle W_{\rm I} \rangle - \langle \Delta F_{\rm I} \rangle_{\rm G}}{\langle Q_\mathrm{h} \rangle} + \frac{k_{\rm B} T_\mathrm{c} \langle \delta \rangle_{\rm G}}{\langle Q_\mathrm{h} \rangle},
\end{split}
\end{equation}
where $ \Delta F_{\rm I}  =  \Delta U_{\rm I}  - T_{\rm c}  \Delta S_{\rm I} $ and $ \Delta S_{\rm I} $ are  fluctuating non-equilibrium free energy and system entropy changes in step I, respectively.
On the other hand, the
stochastic distinguishability $\langle \delta \rangle_{\rm G}$ is a measure of the time-reversal asymmetry through the particle probability densities at gambling times~\cite{manzano2021thermodynamics}. As shown in the Supplemental Material~[SM] the stochastic distinguishability averaged over many realizations of  the GCE takes the form
\begin{equation}
    \langle\delta\rangle_{\rm G} = \displaystyle\left\langle\ln \frac{ {\sigma}_{\taug^\star}^{\star}}{\sigma_{\mathcal{T}} }\right\rangle_{\!\rm G}.\label{eq:SD2}
\end{equation}
Here, $\sigma_{\mathcal{T}}=\left.\sqrt{\langle X^2_t\rangle }\right\vert_{t=\taug}$ 
is the standard deviation of the position in the  BCE 
 yet evaluated at the gambling time~$\mathcal{T}$ extracted from the GCE statistics. Similarly,  $\sigma_{\taug^{\star}}^{\star}
=\left.\sqrt{\langle X^2_{t}\rangle^{\star}} \right\vert_{t=\taug^{\star}}$ 
is the standard deviation of the position in the time-reversed BCE evaluated at the conjugate of the gambling time $\taug ^{\star}=(\tau/4)-\mathcal{T}$.  
In the quasistatic  limit, time reversibility implies that ${\sigma}_{\taug}=\sigma_{\taug^{\star}}^{\star}$, yielding $\langle\delta\rangle_{\rm G}=0$. Moreover, since $\langle W_{\rm I} \rangle \simeq \langle \Delta F_{\rm I} \rangle$ for quasi-static conditions, the two bounds in Eq.~\eqref{eq:etabounds} converge to a single value 
$\Delta \eta \rightarrow (F_{\rm I} - \langle F_{\rm I} \rangle_{\rm G})/(F_{\rm II} - F_{\rm III})$, where $F_{\rm I}^{\rm eq}, F_{\rm II}$ and $F_{\rm III}$ become equilibrium free energies.
Out of equilibrium however, we find ${\sigma}_{\taug^{\star}}^{\star}\leq\sigma_{\mathcal{T}}$ for all values of $\mathcal{T}$, 
which implies that  $\langle\delta\rangle_{\rm G}\leq 0$ for all the cycles times $\tau$ in the GCE. 
The fact that $\langle\delta\rangle_{\rm G}$ is  negative is protocol-specific; for our model it implies that, similarly to the BCE, irreversibilities are detrimental for the efficiency of the GCE, as revealed by our numerical results. Figure~\ref{fig:3}a confirms our theoretical prediction of the sandwich inequality~\eqref{eq:etabounds}, and further reveals that the upper bound in Eq.~\eqref{eq:etabounds} is a bona fide estimate of an upper bound to the efficiency  as it never exceeds one within the explored parameter range in $\tau$. Furthermore, we observe that approaching the quasistatic limit $\Delta \eta $, and thus the lower and upper limits of the sandwich inequality~\eqref{eq:etabounds}, converge to a common value $ T_{\rm c}/T_{\rm h}$ at which $\eta_{\rm G}$ approaches one.

\textit{First-passage statistics and finite-$\tau$ corrections}. 
We now focus on the scaling of the extracted work with the cycle duration $\tau$. 
In the absence of gambling, we refer readers to e.g. Refs.~\cite{sekimoto1997complementarity,schmiedl2007efficiency}, for finite-$\tau$ corrections.
With gambling, a holistic study of the power and efficiency must be done through first passage theory~\cite{redner}.

\begin{figure}[ht]
    \centering
    \includegraphics[width=0.9\columnwidth]{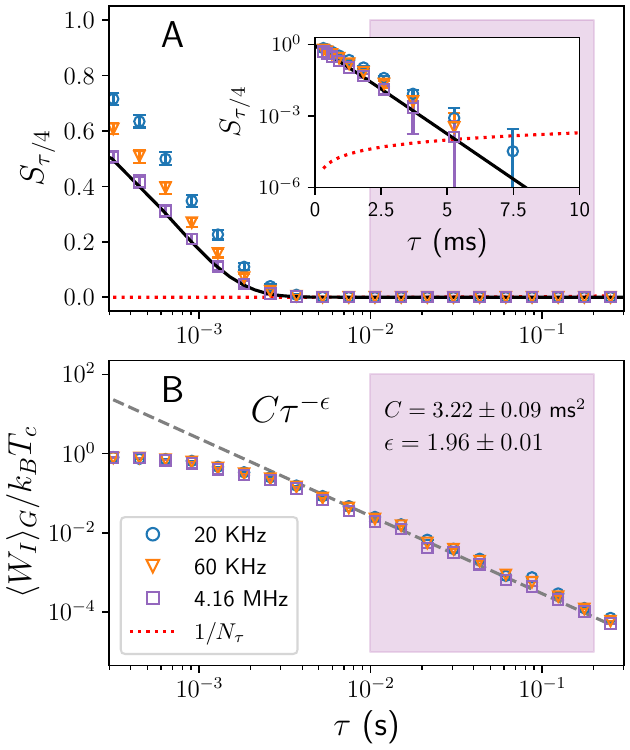}
    \caption{
    Survival probability  $S_{\tau/4}$~(A) for the particle to not cross zero in the interval $[0,\tau/4]$,   and average work exerted by the GCE in step~I $\langle W_{\rm I}\rangle_{\rm G}$~(B) as    a function of cycle duration $\tau$. Each symbol is obtained from $6\times 10^4$ numerical simulations of Eqs.~\eqref{eq:lang}, and~\eqref{eq:kappaC} 
    with a simulation time step $\Delta t = 0.24 \rm{\mu}$s and different sampling frequencies (see legend) up to $(\Delta t)^{-1}=4.16$MHz.  All other  simulation parameter values were set as in Fig.~\ref{fig:3}. 
    In (A)  the red dotted line is $1/N_{\tau} = \tau / \tau_{\rm exp}$  a lower bound to the empirical survival probability detectable in a realistic experiment, assuming that a single experiment lasts $\tau_{\rm exp}=50$s. The  purple shaded areas span cycle times $\tau$ ranging from $10$ms till $200$ ms, i.e. the range used in the BCE experiment~\cite{martinez2016brownian}. In (A), the black solid line is our theoretical prediction for the survival probability in the overdamped limit given by Eq. \eqref{eq:sprob}.
     The inset shows a zoomed view of the figure in log-linear scale. 
    }    
    \label{fig:4}
\end{figure}
Combining the first law of stochastic thermodynamics with  the fact that the energy is conserved  on average over a cycle $\langle \Delta U\rangle_{\rm G} =0$, we can write 
\begin{equation}
    \eta_{\rm G} 
    = 1 + \frac{\langle Q_{\rm c}\rangle_{\rm G}}{\langle Q_{\rm h}\rangle_{\rm G}}\simeq 1 - \frac{\langle W_{\rm I}\rangle_{\rm G}}{\langle Q_{\rm h}\rangle_{\rm G}},
    \label{eq:etaGW}
\end{equation}
where in the second equality in Eq.~\eqref{eq:etaGW} we assumed energy conservation on average along step I, which is a reasonable assumption for any isothermal process. In the quasistatic limit $\langle Q_{\rm h} \rangle_{\rm G}=\langle Q_{\rm h} \rangle=k_{\rm B}T_{\rm h}\ln \sqrt{\kappa_{\tau/2}/\kappa_{\tau'}}$, and $\langle W_{\rm I}\rangle_{\rm G}\leq \langle W_{\rm I}\rangle \mathcal{S}_{\tau/4}$, with
\begin{equation}
    \mathcal{S}_{\tau/4}=1- \int_{0}^{\tau/4}d\taug_0\,\wp(\taug_0),
\end{equation}
the survival probability for the particle to not cross zero in the interval $[0,\tau/4]$. Here,  $\wp(\taug_0)$ denotes the probability density of the first-passage time for the particle position to reach the origin. We derive the  analytical expression for the survival probability $ \mathcal{S}_{\tau/4}$ in the overdamped limit [$m\to 0$ in Eq.~\eqref{eq:lang}]:
\begin{equation}
    \label{eq:sprob}
    \mathcal{S}_{\tau/4} =\dfrac{2}{\pi} \cot^{-1}  \sqrt{2 f_0 \int_{0}^{\tau/4} \exp{2 \left[ f_0 t + 4 \frac{f_{\tau/2} - f_0}{3 \tau^2} t^3\right]} dt} ,
\end{equation}
where $f_0=\kappa_0/\gamma$ and $f_{\tau/2}=\kappa_{\tau/2}/\gamma$ denote respectively the minimum and maximum corner frequencies of the driving protocol. Note that, because $\cot^{-1}(x)\leq \pi/2$ for all $x\geq 0$, we have $\mathcal{S}_{\tau/4}\leq 1$, as expected.
To the authors' knowledge, formula \eqref{eq:sprob} is an original result of this paper (see~\cite{Grebenkov_2015} for known first-passage solutions for Ornstein-Uhlenbeck processes with time-independent coefficients).
Equation~\eqref{eq:sprob} reveals that the survival probability in the overdamped limit decays exponentially fast with the cycle time $\tau$ (see inset in Fig.~\ref{fig:4}A).
Figure~\ref{fig:4}A   shows the survival probability  $  \mathcal{S}_{\tau/4}$  versus the cycle time $\tau$ estimated  from numerical simulations (purple squares) compared with the analytical expression~\eqref{eq:sprob} (black line). 
Notably, even though  the formula~\eqref{eq:sprob} is derived in the overdamped limit, it reproduces our numerical estimates obtained from underdamped Langevin simulations with excellent accuracy. As we show in the SM, the analytical expression~\eqref{eq:sprob} is able to capture the survival statistics of underdamped dynamics for cycle times $\tau\gg\tau_p$, i.e. much larger than the momentum relaxation time, as is the case of the experimental values in many instances (in our case, $\tau_p\sim $ 1 $\rm{\mu}$s and $\tau$ was at least 1 ms). 
In typical experiments however, the sampling frequency of the particle position is of the order of kHz or below, and thus does not allow to track the fine-grained structure of underdamped trajectories. Figure~\ref{fig:4}A shows the effect of undersampling in the estimate of the survival probability, which leads to an overestimation with respect to its actual value. This effect is however only noticeable for fast cycle times $\tau<$ 1 ms i.e. faster than the characteristic corner frequencies  of the GCE.
The observed quick decay of the survival probability to zero with cycle time, together with the inequality $\langle W_{\rm I}\rangle_{\rm G}\leq \langle W_{\rm I}\rangle \mathcal{S}_{\tau/4}$ and Eq.~\eqref{eq:etaGW}  rationalizes the convergence of  $\eta_{\rm G}$ towards one in the quasistatic limit. 

Our analytical progress in first-passage-time statistics also allows us to extract the finite-$\tau$ scaling of power and efficiency of the GCE far beyond the quasistatic limit. In particular, after some algebra, we unveil  
 a scaling law  with the inverse   cycle time squared for the average work in the first step, i.e., 
 \begin{equation}
      \langle W_{\rm I}\rangle_{\rm G} = k_{\rm B}T_{\rm c} (\tau/\tau_w)^{-2} + o(\tau^{-2}),
\label{eq:Wscaling}
\end{equation} 
   with $\tau_w$ a characteristic time that depends on the physical parameters of the GCE (see Supplemental Material [SM]). We test such scaling law  with numerical simulations in Fig.~\ref{fig:4}B where we fit  large-$\tau$ numerical estimates of the average $\langle W_{\rm I}\rangle_{\rm G}/k_{\rm B}T_{\rm c}$ to  $(\tau_w/\tau)^{-\epsilon}$, which yields $\tau_w \simeq \sqrt{3}$ ms and $\epsilon\simeq 1.96$.

\textit{Conclusion.} We have shown that applying feedback control conditioned on first-passage conditions  enable a simultaneous boost of  power and efficiency in Brownian Carnot engines above their classical limits  under realistic experimental conditions, and how such performance boosts depend on potential experimental limitations. It will be interesting in the future to further investigate thermodynamic  trade-offs between the  cost of fast  information acquisition rates and the  energy output of small thermodynamic machines. We expect such ideas to enlighten pioneering experimental devices powered by first-passage strategies~\cite{ribezzi2019large,saha2023information,archambault2024first}.


\

\textit{Acknowledgments.} We acknowledge fruitful discussions with Salamb\^{o}  Dago, Juan MR Parrondo and Ignacio~A.~Mart\'inez. LB and \'ER acknowledge financial support from PNRR MUR project PE0000023-NQSTI. GM acknowledges financial support from the ``Ramón y Cajal'' program (No. RYC2021-031121-I), the CoQuSy project (No. PID2022-140506NB-C21), and the Mar\'ia de Maeztu grant (No. CEX2021-001164-M) funded by MCIU/AEI/10.13039/501100011033 and European Union NextGenerationEU/PRTR.

%

\onecolumngrid
\section*{Supplemental Material}
The Supplemental Material (SM) is organized as follows. In Sec.~\ref{sec:A}, we review theoretical results on stochastic thermodynamics of heat engines with and without feedback, and prove Eq.~\eqref{eq:etabounds} in the Main Text. In Sec.~\ref{sec:B} we provide further details and insights regarding the definition and estimation of the average stochastic distinguishability for the Gambling Carnot Engine (GCE) [Eq.~\eqref{eq:SD2} in the Main Text]. In Sec.~\ref{sec:C}, we provide further details on how our numerical simulations are implemented, together with our estimates of the thermodynamic quantities. In Sec.~\ref{sec:D}, we provide a thorough analytical study for the first-passage-time statistics for the particle to cross the origin in the overdamped limit, and prove the exact analytical formula~\eqref{eq:sprob} in the Main Text. In Sec.~\ref{sec:E}, we show the scaling of the average work in the first stroke of the GCE at large  cycle time [Eq.~\eqref{eq:Wscaling} in the Main Text]. Section~\ref{sec:F} concludes the SM with additional figures that support the numerical results presented in the Main Text. 

\section{Second law and efficiency with and without feedback}
\label{sec:A}

In this section, we first review ground knowledge on stochastic thermodynamics of stochastic heat engines in two scenarios: under deterministic protocols (i.e. in the absence of feedback), and under feedback-control protocols. Next, we briefly review recent results on second laws of thermodynamics at stopping times using martingales, and apply them to derive the bounds for the efficiency given by Eq.~\eqref{eq:etabounds} in the Main Text.

\subsection{Second law and efficiency of stochastic heat engines}

Let us first consider stochastic heat engines where the temperature and stiffness  are controlled externally through deterministic protocols $T_t$ and $\kappa_t$, respectively during time $0\leq t\leq \tau$, with $\tau$ the total cycle time. Within this scenario, the first and second laws of thermodynamics read respectively,
\begin{eqnarray}
    \expval{W}+ \expval{Q}&=&\expval{\Delta U},\label{eq:1BCE}\\
    \langle S_\mathrm{tot} \rangle \equiv \expval{\Delta S}-\int_0^\tau \frac{\expval{\delta Q_t}}{T_t}   &\geq& 0.\label{eq:2BCE}
\end{eqnarray}
Here, $W=\int_0^{\tau}\delta W_t$ and $Q= \int_0^{\tau}\delta Q_t$ denote the stochastic work exerted on the engine and the stochastic heat absorbed by the engine from the baths during a cycle (see Sec. \ref{sec:C} and Eqs.~(\ref{eq:st1}-\ref{eq:st2}) for more detailed definitions). On the other hand, $\Delta U = E(X_t, V_t, \tau) - E(X_t, V_t, 0)$ and $\Delta S = -\ln P(X_t, V_t, \tau) + \ln P(X_0, V_0 ,0)$ denote the system energy and system entropy change during a cycle, where $X_t$ and $V_t$ are the position and velocity of the particle at time $t$, $E(X_t, V_t, t)$ is the energy of the Brownian particle at time $t$, and $P(X_t, V_t, t)$ is its probability density for being at $X_t$ with velocity $V_t$ at time $t$. The brackets $\expval{\cdot}$ above denote averages over many realizations of the cycle and in Eq.~\eqref{eq:2BCE} we introduced the definition of the entropy production.

Assuming that the engine operates cyclically, we have $\expval{\Delta U}=\expval{\Delta S}=0$. On the other hand, for driving protocols such as the BCE, the engine only exchanges non-zero heat on average during the isothermal strokes of the cycle, and hence $\expval{Q}=\expval{Q_{\rm c}}+\expval{Q_{\rm h}}$. As a result, the first and second laws associated with the BCE given by Equations~(\ref{eq:1BCE})-(\ref{eq:2BCE}) take the simplified form
\begin{eqnarray}
    \expval{W}+ \expval{Q_{\rm c}} + \expval{Q_{\rm h}} &=&0, \label{eq:1stlaw}\\
    - \frac{\expval{Q_{\rm c}}}{T_{\rm c}}    -\frac{\expval{ Q_{\rm h}}}{T_{\rm h}}&\geq& 0. 
\end{eqnarray}
The combination of the first and second laws above imply the Carnot bound for the efficiency
\begin{equation}
    \eta = \frac{-\expval{W}}{\expval{Q_{\rm h}}}\leq 1-\frac{T_{\rm c}}{T_{\rm h}}.
\end{equation}
We remind readers that the minus sign in the numerator arises from the fact that the efficiency is defined as the average work expected over a cycle divided by the average heat absorbed from the hot reservoir.

In the presence of generic feedback control, the second law of stochastic thermodynamics can be generalized in various ways. While the first law looks exactly the same as equation \eqref{eq:1stlaw}, the second law is modified by including the entropy production associated with information processing
\begin{eqnarray}
    -\frac{\expval{Q_{\rm c}}}{T_{\rm c}} - \frac{\expval{Q_{\rm h}}}{T_{\rm h}}&\geq& -\langle I \rangle,
\end{eqnarray}
where $\langle I \rangle$ is an entropy exchange representing the information acquired from the system to perform the feedback~\cite{parrondo2015thermodynamics}. The specific expression of the information-theoretic quantity $\langle I \rangle$ may vary depending on the type of measurement and feedback control applied and several have been proposed in the literature~\cite{parrondo2015thermodynamics,Lutz15,Koski15,Camati16,Paneru18,Paneru2020,Saha21,Fadler23,Saha23}. In any case, a generalized efficiency might be defined that verifies
\begin{equation}
    -\frac{\langle W\rangle }{\langle Q_{\rm h}\rangle +k_{\rm B}T_{\rm c} \langle I\rangle/\eta_C}\leq \eta_C,
\end{equation}
that is, one retrieves the Carnot bound when incorporating the entropy flow due to feedback control. Moreover, depending on the value of $\expval{I}$, the ratio $-\expval{W}/\expval{Q_{\rm h}}$ may even exceed the Carnot limit.

\subsection {Second law at gambling times and efficiency of stochastic heat engines using gambling}

On the other hand, the second law can be also generalized for processes that incorporate gambling at stopping times~\cite{manzano2021thermodynamics}. In that case the total entropy production until a (bounded) stopping condition $\taug \leq \tau$ is met verifies 
\begin{eqnarray} \label{eq:secondlawstoppings}
    \langle S_\mathrm{tot} \rangle_{\rm G} = \langle \Delta S \rangle_{\rm G} -\frac{\expval{Q_{\rm c}}_{\rm G}}{T_{\rm c}} - \frac{\expval{Q_{\rm h}}_{\rm G}}{T_{\rm h}}&\geq& -k_{\rm B} \expval{\delta}_{\rm G},
\end{eqnarray}
where all averages denoted as $\langle \cdot \rangle_{\rm G}$ are evaluated at the (stochastic) stopping time $\mathcal{T}$ over many realizations. Here above we introduced the stochastic distinguishability evaluated at stopping times:
\begin{eqnarray}
    \delta =  \left.\ln \frac{P(x,v,t)}{P^{\star}(x,-v,t^\star)}\right\vert_{x=X_\taug, v=V_\taug,t=\mathcal{T},t^\star=(\tau/4)-t},
\end{eqnarray}
which measures the time-asymmetry of the process through the comparison of the particle density at the stopping time, $P(x,v,t=\taug)$ with its time-reversed counterpart $P^{\star}(x,-v,t^{\star}=t-\taug)$ (see \ref{sec:B} for more details). 

The above inequality \eqref{eq:secondlawstoppings} follows from the fact that $\exp\left(-S_\mathrm{tot} - k_{\rm B} \delta\right)$ is a martingale, from which the fluctuation theorem at stopping times follows~\cite{manzano2021thermodynamics}:
\begin{equation}
    \langle \exp\left(-S_\mathrm{tot} - k_{\rm B} \delta\right)\rangle_{\rm G} = 1,
\end{equation}
and applying Jensen's inequality to the equation above directly leads to the second law inequality~\eqref{eq:secondlawstoppings}.
In the GCE we are interested in applying gambling only over the first isothermal compression stroke of the cycle, using as stopping time the gambling time in Eq.(5) of the main text. Particularizing the second law in Eq.~\eqref{eq:secondlawstoppings} for this stroke only we obtain:
\begin{eqnarray} \label{eq:secondGCE}
    \langle W_{\rm I} \rangle_{\rm G} - \langle \Delta F_{\rm I} \rangle_{\rm G} \geq - k_B T_{\rm c} \langle \delta \rangle_{\rm G},
\end{eqnarray}
with $W_{\rm I}$ the work exerted to the Brownian particle during step I, and $\Delta F_{\rm I} = \Delta U_{\rm I} - k_B T_{\rm c} \Delta S_{\rm I}$ the corresponding nonequilibrium free-energy change, all of which are again evaluated at the stopping time $\taug$.


Neglecting further contributions to the work due to the  quench at $\mathcal{T}$, we have that the work exerted at the end of the first stroke of the GCE is just the work exerted until the stopping time. As a consequence, the efficiency of the engine reads
        \begin{equation} \label{eq:GCE}
            \eta_{\rm G} = -\frac{ \langle W \rangle_{\rm G}}{\langle Q_\mathrm{h} \rangle_{\rm G}}\simeq \frac{- \langle W_{\rm I} \rangle_{\rm G} - \langle W_{\rm II} \rangle - \langle W_{\rm III} \rangle-  \langle W_{\rm IV} \rangle}{\langle Q_\mathrm{h} \rangle},
        \end{equation}
        where $\langle W_{\rm I} \rangle$ for $i=$II, III, IV is the work exerted in the other three strokes and $\langle Q_\mathrm{h}\rangle$ is the heat absorbed from the hot bath during the entire cycle. 
Here above the second equality is an approximation considering that during the remaining strokes of the GCE, where the same driving protocol than the original BCE applies, work and heat contributions equal those in the BCE. 

We now rewrite Eq.~\eqref{eq:GCE} by adding and subtracting in the numerator the average work $\langle W_{\rm I} \rangle$ exerted on the particle in the absence of gambling,
        \begin{align} \label{eq:GCEsplit}
            \eta_{\rm G} =  \underbrace{\frac{- \langle W_{\rm I} \rangle - \langle W_{\rm II} \rangle - \langle W_{\rm III} \rangle-  \langle W_{\rm IV} \rangle}{\langle Q_\mathrm{h} \rangle}}_{\displaystyle\eta} +  \underbrace{\frac{\langle W_{\rm I} \rangle - \langle W_{\rm I} \rangle_{\rm G}}{\langle Q_\mathrm{h} \rangle}}_{\displaystyle\Delta\eta}
           ,
        \end{align}
where we identified the efficiency of the standard Carnot engine without gamgling $\eta$ (first term) and an extra contribution $\Delta \eta$ (second term) due to gambling in the compression stroke.       Equation~\eqref{eq:GCEsplit} suggests that a reduction on the work exerted in the isothermal compression may be accompanied by an enhancement of the efficiency.

We can establish an upper bound for $\Delta \eta$ based on the second law at stopping times as applied to the first stroke of the GCE in Eq.~\eqref{eq:secondGCE} by introducing it as a bound for $\langle W_{\rm I} \rangle_{\rm G}$ in the defition of $\Delta \eta$. That give us:
        \begin{equation}
            \Delta \eta \leq \frac{\langle W_{\rm I} \rangle - \langle \Delta F_{\rm I} \rangle_{\rm G}}{\langle Q_\mathrm{h} \rangle} + \frac{k_{\rm B} T_\mathrm{c} \langle \delta \rangle_{\rm G}}{\langle Q_\mathrm{h} \rangle}.
            \label{eq:ub}
        \end{equation}
On the other hand, we also obtain a lower bound  for $\Delta\eta$, using the fact that the quantity $W_{\rm I}-\Delta F_{\rm I}-k_{\rm B}T_{\rm c}\delta$ is a submartingale, which is a direct consequence of $\exp(-S_\mathrm{tot} -k_B \delta)$ being a martingale~\cite{manzano2021thermodynamics}. For any submartingale $A$ and two ordered stopping times, such that $\taug_1 < \taug_2$, it follows that $\langle A \rangle_{\taug_1} \geq \langle A \rangle_{\taug_2}$ (see e.g. Eq.~(5) in~\cite{manzano2024thermodynamics}). Therefore, by taking $A = W_{\rm I}-\Delta F_{\rm I}-k_{\rm B}T_{\rm c}\delta$, $\taug_1 = \taug$ the gambling time of the GCE and $\taug_2 = \tau/4$, one has that:
        \begin{equation}
            \langle W_{\rm I} \rangle_{\rm G} - \langle \Delta F_{\rm I} \rangle_{\rm G} + k_{\rm B} T_\mathrm{c}\langle \delta \rangle_{\rm G} \leq  \langle W_{\rm I} \rangle - \langle \Delta F_{\rm I} \rangle,
            \label{eq:submart}
        \end{equation}
which is a special case of Eq.~(5) in~\cite{manzano2024thermodynamics} applied to the ordered stopping times $\mathcal{T}$ and $\tau/4$. We recall that the gambling time $\mathcal{T} \leq \tau/4$ with probability one as follows from its definition in Eq.(5) of the main text. In the above inequality we also used the fact that the stochastic distinguishability becomes zero when evaluated at the deadline time.

The inequality~\eqref{eq:submart} can then be directly used to obtain the following lower bound on $\Delta\eta$:
        \begin{equation}
            \Delta \eta \geq \frac{\langle \Delta F_{\rm I} \rangle - \langle \Delta F_{\rm I} \rangle_{\rm G}}{\langle Q_\mathrm{h} \rangle} + \frac{k_{\rm B} T_\mathrm{c} \langle \delta \rangle_{\rm G}}{\langle Q_\mathrm{h} \rangle}.
            \label{eq:lb}
        \end{equation}

Combining the inequalities~\eqref{eq:lb} and~\eqref{eq:ub}, we obtain the sandwich inequality given by Eq.~\eqref{eq:etabounds} in the main text:
        \begin{equation}
            \frac{\langle \Delta F_{\rm I} \rangle - \langle \Delta F_{\rm I} \rangle_{\rm G}}{\langle Q_\mathrm{h} \rangle} + \frac{k_{\rm B} T_\mathrm{c} \langle \delta \rangle_{\rm G}}{\langle Q_\mathrm{h} \rangle} \leq \Delta \eta \leq \frac{\langle W_{\rm I} \rangle - \langle \Delta F_{\rm I} \rangle_{\rm G}}{\langle Q_\mathrm{h} \rangle} + \frac{k_{\rm B} T_\mathrm{c} \langle \delta \rangle_{\rm G}}{\langle Q_\mathrm{h} \rangle}.
        \end{equation}
        
As mentioned  in the main text,  $\langle \delta \rangle_{\rm G} \rightarrow 0$ in the quasistatic limit because the probability densities of the Brownian particle during the forward and time-reversed protocols both approach the instantaneous equilibrium distribution at all times during the cycle. Moreover, reversibility conditions enforced by the quasi-static condition imply no entropy production during the strokes and hence in particular $\langle W_{\rm I} \rangle \rightarrow \langle \Delta F_{\rm I} \rangle$ and $\langle Q_{\rm h} \rangle = - \langle \Delta F_{\rm III} \rangle$, minus the change in free energy in the third stroke. As a consequence the two bounds in Eq.~\eqref{eq:etabounds} of the main text converge to a single quantity and the improvement in efficiency of the GCE becomes:
\begin{eqnarray} \label{eq:quasi-static}
    \Delta \eta \rightarrow \frac{\langle F_{\rm I} \rangle - \langle F_{\rm I} \rangle_{\rm G}}{\langle F_{\rm II} \rangle - \langle F_{\rm III} \rangle} = \frac{T_{\rm c}}{T_{\rm h}} \left( \frac{\ln \kappa_{\tau/4} - \langle \ln\kappa_{\mathcal{T}}\rangle_{\rm G}}{\ln \kappa_{\tau/4} - \ln \kappa_{0}} \right)
\end{eqnarray}
where in the second equality we have used that the equilibrium free energies $F_{\rm I} = - \ln Z_{\rm I} = \frac{k T_{\rm I}}{2}\ln \kappa_{\rm I}$ for $i= \rm I, \rm II, \rm III, \rm IV$ with $\kappa_i = \{\kappa_{\tau/4}, \kappa_{\tau/2}, \kappa_{3\tau/4}, \kappa_0\}$ and $T_i = \{ T_{\rm c}, T_{\rm c}, T_{\rm h}, T_{\rm h} \}$, and the micro-adiabaticity condition, $\frac{\kappa_{\tau/2}}{\kappa_\tau/4}= \frac{\kappa_{3\tau/4}}{\kappa_0}$.
The above equation implies that whenever $\langle \ln \kappa_\taug \rangle_{\rm G} \rightarrow \ln \kappa_0$ in the quasi-static limit (the gambling condition is reached quickly), the expression inside the parenthesis become $1$, the efficiency improvement $\Delta \eta \rightarrow T_{\rm c}/T_{\rm h}$ and hence the efficiency of the GCE becomes one, $\eta_{\rm G} \rightarrow 1$.

\section{Stochastic distinguishability}
\label{sec:B}

In this Section, we put at work the concept of stochastic distinguishability applied to the cold isothermal compression in the GCE, i.e. for $0\leq t\leq \tau/4$. To this aim, let us first consider the underdamped Langevin equation
 [Eq.~\eqref{eq:lang} in the Main Text] associated with the BCE along the cold isothermal
\begin{equation}
    m\ddot{X}_t = - \gamma \dot{X}_t - \kappa_t X_t + \sqrt{2\gamma k_{\rm B}T_c}\,\xi_t,
    \label{eq:lang2a}
\end{equation}
with $\kappa_t$ and $T_t$ following the {\em deterministic} protocols from the BCE. 
The probability density $P(x,v,t)\equiv \mathrm{Prob} \left\lbrace X_t=x, V_t=v | (X_0,V_0)\sim P_0(x,v) \right\rbrace$ to find the GCE in the position $x$ and velocity $v$ given that its initial condition was of equilibrium $P_0(x,v)=\exp(-\kappa_0 x^2/2k_{\rm B}T_c)\exp(-m v^2/2k_{\rm B}T_c)$ follows, at times $t>0$, the Kramers-Klein-Chandrasekhar equation
\begin{equation}
    \frac{\partial}{\partial t}P(x,v,t)= \left(- v \frac{\partial}{\partial x} + \frac{\kappa_t x}{m} \frac{\partial}{\partial v} +\frac{\gamma k_{\rm B}T_c}{m}\frac{\partial^2}{\partial v^2}+\frac{\gamma}{m}\frac{\partial}{\partial v} v\right) P(x,v,t) .\label{eq:KC1}
\end{equation}
Analogously, we will consider also a virtual, time-reversed process (with superscript $^{\star}$), whose initial state is drawn with  the statistics of the BCE at the end of the first isothermal, but including a sign reversal of the velocity, i.e. $P^{\star}_0(x,v)=P_{\tau/4}(x,-v)$. The evolution of the time-reversed process at times $t>0$ follows the underdamped SDE
\begin{equation}
    m\ddot{X}_t = - \gamma \dot{X}_t - \kappa_t^{\star} X_t + \sqrt{2\gamma k_{\rm B}T_c}\,\xi_t,
    \label{eq:langr}
\end{equation}
where $\kappa_t^{\star}=\kappa_{\tau/4-t}$  follows the BCE step I but ``run in reverse" i.e. time-reversed with respect to the reference horizon time $\tau/4$. The probability density $P^{\star}(x,v,t)$ to be at $x$ and $v$ at time $t<\tau/4$ in the time-reversed process obeys the Kramers-Klein-Chandrasekhar equation
\begin{equation}
    \frac{\partial}{\partial t}P^{\star}(x,v,t)= \left(- v \frac{\partial}{\partial x} + \frac{\kappa_t^\star x}{m} \frac{\partial}{\partial v} +\frac{\gamma k_{\rm B}T_c}{m}\frac{\partial^2}{\partial v^2}+\frac{\gamma}{m}\frac{\partial}{\partial v} v\right) P^{\star}(x,v,t) .
    \label{eq:KC2}
\end{equation}
with initial condition $P^{\star}(x,v,0)=P(x,-v,\tau/4)$. \\
The average stochastic distinguishability associated with the GCE is given by
\begin{equation}
    \left\langle\delta\right\rangle_{\rm G} =  \left\langle\left.\ln \frac{P(x,v,t)}{P^{\star}(x,-v,t^\star)}\right\vert_{x=X_\taug, v=V_\taug,t=\mathcal{T},t^\star=(\tau/4)-t}\right\rangle_{\rm G} .
    \label{eq:stopgg}
\end{equation}
Here, the densities $P$ and $P^{\star}$ are the solutions of the Kramers-Klein-Chandrasekhar equations~\eqref{eq:KC1} and~\eqref{eq:KC2}, respectively, i.e. those associated with the process without stopping conditions. On the other hand, the average in Eq.~\eqref{eq:stopgg} is obtained by evaluating the times  at $t=\mathcal{T}$ and $t^\star=(\tau/4)-\mathcal{T}$, where $\mathcal{T}=\min(\mathcal{T}_0,\tau/4)$ [see Eq.~\eqref{stop_time} in the main text] is stochastic and drawn from the GCE realizations with feedback control. Such mixture of statistics makes the calculation of the average stochastic distinguishability an analytical quest, therefore we approached the problem by combining theory and numerical simulations. \\
In our simulations, we used cycle times larger than the momentum relaxation time, i.e. $\tau\gg \tau_p=m/\gamma$. In this setting the velocity has sufficient time to relax to equilibrium and thus we can rely on the approximation
\begin{align}
\left\langle\delta\right\rangle_{\rm G} &\simeq  
\left\langle\left.\ln \frac{P(x,t)}{P^{\star}(x,t^\star)}\right\vert_{x=X_\taug, t=\mathcal{T},t^\star=(\tau/4)-t}\right\rangle_{\rm G} .
    \label{delta_T0}
\end{align}
where $P(x,t)$ and $P^{\star}(x,t^\star)$ are the solutions of the Fokker-Planck equation associated with the forward and backward overdamped Langevin equations ($m=0$ in Eqs.~\eqref{eq:lang2a} and~\eqref{eq:langr}),  respectively. More precisely,  $P(x,t)\equiv \mathrm{Prob} \left\lbrace X_t=x | X_0\sim P_0(x_0) \right\rbrace$ is the solution of the Fokker-Planck equation
\begin{equation}
    \frac{\partial}{\partial t}P(x,t) = -\frac{\kappa_t}{\gamma} \frac{\partial}{\partial x} (x P(x,t)) + \frac{k_{\rm B}T_c}{\gamma }\frac{\partial^2}{\partial x^2}  P(x,t),
    \label{eq:FPE1}
\end{equation}
with initial (equilibrium) condition  $P_0(x)=\exp(-\kappa_0 x^2/2k_{\rm B}T)$. On the other hand, $P^{\star}(x,t)$ solves the Fokker-Planck equation
\begin{equation}
    \frac{\partial}{\partial t}P^{\star}(x,t) = -\frac{\kappa_t^{\star}}{\gamma} \frac{\partial}{\partial x} (x P^{\star}(x,t)) + \frac{k_{\rm B}T_c}{\gamma }\frac{\partial^2}{\partial x^2}  P^{\star}(x,t),\label{eq:FPE2}
\end{equation}
with initial condition $P_0^{\star}(x)=P(x, \tau/4)$. It has been shown that (see e.g.~\cite{martinez2016engineered}), for initial Gaussian distributions, the solutions of Eqs.~(\ref{eq:FPE1}-\ref{eq:FPE2}) are Gaussian at all times,  even when the particle is driven far from equilibrium by varying its stiffness. Hence, the variance of the position  $\sigma_t^2 = \la X_t^2 \ra$ and $(\sigma_t^{\star})^2 = \la X^{\star 2}_t \ra$ fully characterize the probability densities $P(x,t)$ and $P^{\star}(x,t)$. More precisely, one can show that
\begin{equation}
    P(x,t) = \sqrt{\frac{1}{2\pi\sigma^2_t}}\exp\left(-\frac{ x^2}{2\sigma^2_t}\right),\qquad  P^{\star}(x,t) = \sqrt{\frac{1}{2\pi(\sigma_t^{\star})^2}}\exp\left(-\frac{ x^2}{2(\sigma_t^{\star})^2}\right),
\end{equation}
where the variances follow the following first-order ordinary differential equations with time-dependent coefficients
\begin{align}
    \frac{d}{dt}\sigma^2_t &= 2D - 2 f_t \sigma^2_t, \\
   \frac{d}{dt}(\sigma^{2}_t)^{\star}  &= 2D - 2f_t^{\star} (\sigma^{2}_t)^{\star},
\end{align}
with $f_t=\kappa_t/\gamma$, $f_t^{\star}= (\kappa_{(\tau/4)-t})/\gamma$, and $D=k_{B}T_c/\gamma$. 
Knowledge of the variance at times $t\leq \tau/4$ allows to write a concise expression of the stochastic distinguishability at stopping times.
\begin{align}
\left\langle\delta\right\rangle &\simeq  \left\langle\left.\ln \frac{P(x,t)}{P^{\star}(x,t^\star)}\right\vert_{x=X_\taug, t=\mathcal{T},t^\star=(\tau/4)-t}\right\rangle , \\
    & = \left\langle\left.\left(\frac{1}{2}\ln \frac{(\sigma^2_{(\tau/4)-\mc{T}})^{\star}}{\sigma^2_{\mc{T}}}
    -\frac{1}{2} \left[\frac{1}{\sigma^2_{\mc{T}}} - \frac{1}{(\sigma^2_{(\tau/4)-\mc{T}})^{\star}}\right]X_{\mc{T}}^2\right)\right\vert \mathcal{T}<\tau/4\right\rangle (1-\mathcal{S}_{\tau/4}), \\
    & \simeq  \frac{1}{2}\left\langle\ln \frac{(\sigma^2_{(\tau/4)-\mc{T}})^{\star}}{\sigma^2_{\mc{T}}}\right\rangle,
    \label{delta_T}
\end{align}
where in the second line $\mathcal{S}_{\tau/4}$ denotes the survival probability.
We have used the fact that $X_{\mc{T}} = 0$, with probability one. 
Further, rather than integrating with respect to the density of stopping times (see appendix) we instead approximate $\langle \delta \rangle$ with an empirical average from $N$ samples of the stopping time, i.e. 
\begin{equation}
\langle\delta_\taug\rangle
\simeq \frac{1}{2N}\sum_{i=1}^{N_{\rm g}} \ln\left[\frac{(\sigma^2_{(\tau/4)-\mathcal{T}_i})^{\star}}{\sigma^2_{\mathcal{T}_i}}\right] ,  
\end{equation}
with $\mathcal{T}_i$ the $i-$th value of the stopping time obtained in simulations, $N$ the total number of cycles, and $N_{\rm g}\leq N$ the number of gambles, i.e. the number of times in which $\mathcal{T}_i<\tau/4$.

\section{Simulation details}
\label{sec:C}

In this section, we provide an in-depth description on how all our numerical simulations are implemented.  To this aim, 
 we first rewrite the stochastic differential equation~\eqref{eq:lang} associated with the dynamics of the particle as
 \begin{equation}
    \tau_p\frac{d^2}{dt^2}{X}_t = -  \frac{d}{dt}X_t - \frac{\kappa_t}{\gamma} X_t + \sqrt{2 \frac{k_{\rm B}T_t}{\gamma}}\,\xi_t,
    \label{eq:lang2}
\end{equation}
where
\begin{equation}
    \tau_p = \frac{m}{\gamma},
    \label{eq:tau_p}
\end{equation}
denotes the momentum relaxation time. Let us now introduce new dimensionless coordinates $s,Y$
\begin{equation}
    s=\frac{t}{\tau_p},\qquad Y=\frac{X}{v_c\tau_p},
\end{equation}
with
\begin{equation}
    v_c=\sqrt{\frac{2k_{\rm B}T_{\rm c}}{m}},
<c \end{equation}
a characteristic velocity. Applying this coordinate change to Eq.~\eqref{eq:lang2} we obtain the following second-order SDE in dimensionless form
\begin{equation}
    \frac{d^2}{ds^2}Y_{s} = -\frac{d}{ds}Y_{s} - \omega_{s} Y_{s} + \sqrt{{\mathsf{T}}_s}\,\xi_s,
    \label{eq:langN}
\end{equation}
where in Eq.~\eqref{eq:langN} we have introduced the dimensionless quantities
\begin{equation}
\label{eq:rescaled_omega_T}
    \omega_s =\frac{\tau_p}{\gamma} \kappa_s,\quad \mathsf{T}_s=\frac{T_s}{T_{\rm c}}.
\end{equation}
In our numerical simulations, we numerically integrated the dimensionless stochastic differential equation~\eqref{eq:langN} for the stochastic protocol defined in the main text, 
and later on convert all the dimensionless quantities into their  analogues with physical dimensions. In practice, when working with the dimensionless stochastic differential equation~\eqref{eq:langN}, we use the dimensionless protocol
\begin{equation}
    \omega_s =\frac{\tau_p}{\gamma} \kappa_s = \begin{cases}
        \omega_0 + \zeta s^2&  0\leq s < \taug \qquad\; \text{(Ia)} \\
        \omega_0 + \zeta (\tau/4)^2 & \taug\leq s < \tau/4 \quad\; \text{(Ib)}\\
        \omega_0 + \zeta s^2 & \tau/4\leq s< \tau/2 \quad \text{(II)}\\
        \omega_0 + \zeta (\tau - s)^2  & \tau/2\leq s \leq  \tau  \quad \text{(III) and (IV)}
    \end{cases} 
\label{eq:kappaC2}
\end{equation}
for the dimensionless stiffness, and 
\begin{align}
    \mathsf{T}_s = \frac{T_s}{T_{\rm c}}= \begin{cases}
        1& 0 \leq s < \displaystyle\frac{\tau}{4\tau_p}\qquad \text{(I)} \\
        \sqrt{\displaystyle\frac{\omega_0 + \zeta s^2}{\omega_{\tau/4}}}  & \displaystyle\frac{\tau}{4\tau_p} \leq s <\displaystyle\frac{\tau}{2\tau_p} \quad \text{(II)} \\
        \displaystyle\frac{T_{\rm h}}{T_{\rm c}}&\displaystyle\frac{\tau}{2\tau_p}  \leq s <\displaystyle\frac{\tau^\star}{\tau_p} \;\;\quad \text{(III)} \\
        \sqrt{\displaystyle\frac{\omega_0 + \zeta \left(\displaystyle\frac{\tau}{\tau_p}-s\right)^2}{\omega_0}}& \displaystyle\frac{\tau^\star}{\tau_p} \leq s <\displaystyle\frac{\tau}{\tau_p}, \;\;\qquad \text{(IV)} 
    \end{cases}
    \label{eq:TC2}
\end{align}
for the dimensionless temperature, 
where,
\begin{equation}
    \zeta=4(\omega_{\tau/2}-\omega_{0})\left(\frac{\tau_p}{\tau}\right)^2.
\end{equation}

In our simulations, we integrate Eq.~\eqref{eq:langN} numerically using the Euler-Mayurama algorithm by discretizing time in fixed steps $\Delta t$.  In all our simulations (except for Fig.~\ref{fig:3}) we set $\Delta t = 0.2 \mu$s to be less than the momentum relaxation time $\tau_p = 0.25\mu$s (see caption of Fig. \ref{fig:3} for experimental parameters).  This leads to a dimensionless time step of $\Delta s=\Delta t/\tau_p=0.8$. The discrete-time evolution of the dimensionless position $Y_t$ and velocity $R_t = V_t / v_c$ read
\begin{align}
    Y_{s + \Delta s} &= Y_s + R_s\Delta s, \label{eq:em_x} \\ 
    R_{s + \Delta s} &= R_s - R_s \Delta s - \omega_s Y_s\Delta s
    + \sqrt{\mathsf{T}\Delta s} \, \eta_t,
    \label{eq:em_v}
\end{align}
where  $\eta_t\sim \mathcal{N}(0,1) $ is sampled from a normal distribution with zero mean  and unit variance. The initial position and velocity are sampled from their equilibrium (Maxwell-Boltzmann) distributions,
\begin{equation}
    X_0 \sim \mathcal{N}\left(0, \frac{k_BT_{\rm c}}{\kappa_0}\right) \rightarrow Y_0 \sim \mathcal{N}\left(0, \frac{k_BT_{\rm c}}{\kappa_0v_c \tau_p}\right), 
    \end{equation}
    for the position, and
    \begin{equation}
    V_0 \sim \mathcal{N}\left(0, \frac{k_BT_{\rm c}}{m}\right) \rightarrow R_0 \sim \mathcal{N}\left(0, \frac{k_BT_{\rm c}}{mv_c}\right),    
    \end{equation}
    for the velocity,     where here $\mathcal{N}(\mu,\sigma^2)$ denotes a Gaussian distribution with mean $\mu$ and variance $\sigma^2$. 

To model the effects of the experimental sampling rate on the performance of the GCE (see figure \ref{fig:efficiency-sampling}), we sub-sample points obtained from \eqref{eq:em_x} and \eqref{eq:em_v} at regular intervals $\Delta r = k\Delta t$, where $k$ is an integer subsampling factor. The stopping time $\mathcal{T}$ is then computed as the first time step $i\Delta r$ at which $Y_{(i+1)\Delta r}Y_{i\Delta r} < 0$, indicating a change of sign. After solving equations \eqref{eq:em_x} and \eqref{eq:em_v}, sub-sampling the solutions, and rescaling them back to natural units, we obtain an ensemble of trajectories for position and velocity. We denote them respectively as $\textbf{X} = \{X_{t_i} \}$ and $\textbf{V} = \{V_{t_i} \}$, with $ t_i \in \{0, \Delta r, \dots, n\Delta r \}$.
The increments of thermodynamic quantities in $[t,t+\Delta t]$ were computed from the stochastic trajectories as follows:
\begin{eqnarray}
    \delta W_t &=& \frac{1}{2}\big ( \kappa_{t + \Delta r}X_t^2 - \kappa_{t}X_t^2\big ) \quad \text{Work}\label{eq:st1} \\
    \delta Q_t &=& \frac{1}{2}\big ( \kappa_{t + \Delta r}X_{t + \Delta r}^2 - \kappa_{t + \Delta r}X_t^2\big ) + \frac{m}{2}\big (V_{t + \Delta r}^2  - V_{t}^2\big ) \quad \text{Heat} \\
    d E_t &=& \delta W_t+\delta Q_t \quad \text{Energy} \\
    d S_t &=&  \frac{k_B}{2}\Big (\ln \frac{\sigma^2_{t+\Delta r}}{\sigma^2_t} + \frac{X_{t + \Delta r}^2}{\sigma^2_{t+\Delta r}} - \frac{X_t^2}{\sigma^2_t} \Big ) \quad \text{Non-equilibrium entropy} \\
    d F_t &=& d E_t - T_t d S_t \quad \text{Non-equilibrium free energy}\label{eq:st2}
\end{eqnarray}


\section{First-passage-time calculations}
 \label{sec:D}

We prove here the formula for the survival probability we used in the main text. We briefly outline here the scheme for the proof: i) obtain the solution of the Fokker-Planck equation (FPE) in the case of open boundary conditions; ii) obtain the solution of the propagator in the case of absorbing boundary conditions using the image method; iii) obtain the formula for the survival probability by integrating over the appropriate boundary the propagator of point (ii). To make this part more comprehensible, we prefer to split it into three paragraphs.

As a preliminary, here we write the dimensionless Langevin equation in the overdamped limit, which we use in subsequent appendices to calculate survival probabilities.
\begin{equation}
\label{eq:langOD}
    \dfrac{d}{du} Z_u = -\mathsf{K}_u Z_u + \sqrt{\mathsf{T}_u} \xi_u
\end{equation}
Introducing a space-time coordinate change as in \eqref{eq:langN}, imposing that both $u$ and $Z$ are dimensionless, leads to the following rescaling:
\begin{equation}
\label{eq:od_scaling}
   u = \frac{t}{\gamma/\kappa_0}, \quad \quad  \quad Z_u = \frac{X_u}{\sqrt{\frac{2k_BT_{\rm c}}{\kappa_0}}},
   \quad \quad \quad \mathsf{K}_u = \frac{\kappa_u}{\kappa_0}, \quad \quad \quad \mathsf{T}_u = \frac{T_u}{T_{\rm c}}.
\end{equation}
\subsection{Solution of the FPE with open boundary conditions}
Working with dimensionless quantities, let us denote with ${P(z,u|z_0,u_0) \equiv \mathrm{Prob} \left\lbrace Z_u=z | Z_{u_0}=z_0 \right\rbrace}$. In this section, we use the same letter $P$ for the propagator, but, for a better understading, we prefer specifying the initial conditions $z_0,u_0$ in the arguments of the function.
From the Langevin equation, we can infer the Fokker-Planck equation (FPE) for this propagator
\begin{equation}
    \label{eqn:FPE}
    \dfrac{\partial P(z,u|z_0, u_0)}{\partial u} = \dfrac{\mathsf{T}_u}{2} \dfrac{\partial^2 P(z,u|z_0, u_0)}{\partial z^2} + \mathsf{K}_u \dfrac{\partial P(z,u|z_0, u_0)}{\partial z} ,
\end{equation}
with initial condition $P(z,u_0|z_0, u_0)= \delta(z-z_0)$. In the case of open boundary conditions, i.e. $\lim_{z\to \pm \infty} P(z,u|z_0, u_0)=0$,
the solution to this parabolic partial differential equation is known~\cite{polyanin}. It reads
\begin{equation}
    \label{eqn:propagator}
    P(z,u|z_0, u_0) = \dfrac{\exp \left( \psi_{u,u_0} \right)}{\sqrt{2 \pi \varphi_{u,u_0}}} \exp\left(- \dfrac{(z e^{\psi_{u,u_0}}-z_0)^2}{2 \varphi_{u,u_0}} \right)
\end{equation}
where 
\begin{equation}
    \psi_{u,u_0} \equiv \int_{u_0}^{u} \mathsf{K}_r dr,
    \label{eq:psi}
\end{equation}
and
\begin{equation}
    \varphi_{u,u_0} \equiv \int_{u_0}^{u} \mathsf{T}_r \exp \left(2\int_{u_0}^{r} \mathsf{K}_{r'} dr' \right) dr = \int_{u_0}^{u} \mathsf{T}_r e^{2\psi_r} dr,
    \label{eq:phi}
\end{equation}
are two deterministic functions of time that depend explicitly on the protocol through $\mathsf{K}_u$ and $\mathsf{T}_u$. As the process is not time-homogeneous, we highlighted the dependence on the initial time $u_0$. 

\subsection{Solution of the FPE with one absorbing boundary}
In this paragraph we consider $u_0=0$ and we drop the dependence on $u_0$ everywhere.
Let us consider here the case in which there is an absorbing boundary at $z=0$, i.e. $P_a(0,u|z_0)=0$, here we use the subscript $a$ to make a distinction with the open boundaries case. To fix the ideas, let us consider that $z_0>0$, while the case $z_0<0$ is completely analogous. It is possible to obtain the propagator $P_a(z,u|z_0)$ using the so-called image method~\cite{redner}. The key idea is that, since the FPE~\eqref{eqn:FPE} is a linear differential equation, a linear combination of solutions is still a solution of Eq.~\eqref{eqn:FPE}. The proper linear combination of solutions of the open boundary problem  fulfilling the boundary condition is the following
\begin{equation}
    \label{eqn:P_a}
    P_a(z,u|z_0) = P(z,u|z_0) - P(z,u|-z_0),
\end{equation}
where the functional form of $P(z,u|z_0)$ is given in eq \eqref{eqn:propagator}, and the result is valid only for $z\in [0,+\infty[$.
Physically speaking, this can be seen as placing a negative source of probability originating from the symmetric point of $z_0$ with respect to the origin. In conclusion, $P_a(z,u|z_0)$ solves Eq.~\eqref{eqn:FPE} and vanishes at $z=0$.
\subsection{Survival probability in dimensionless units} The probability distribution $P_a(z,u|z_0)$   is not normalized, thus it is often called `defective probability' in the literature. Indeed, since there is an absorbing boundary, many of the stochastic trajectories starting in $z_0$ get absorbed within time $u$. To obtain the probability that a trajectory is not absorbed, i.e. survives, up to time $u$ we need to integrate the propagator in $z$ over the interval $[0,+\infty)$. This probability is called, indeed, {\em survival probability} and reads, for the case $z_0>0$
\begin{equation}
    \mathcal{S}_u(z_0) = \int_{0}^{+\infty} P_a(z,u|z_0) dz = \erf \left(\dfrac{z_0}{\sqrt{2\varphi_{u}}} \right),
\end{equation}
where to obtain the last result we simply integrated eq \eqref{eqn:P_a}.
By substituting the explicit form of the function $\varphi_{u}$ we get
\begin{equation}
    \mathcal{S}_u(z_0) = \erf \left( \dfrac{z_0}{\sqrt{2 \int_{0}^{u} \mathsf{T}_r e^{2\psi_r}dr}} \right),
    \label{eq:sprob_resc}
\end{equation}
which holds  for $z_0>0$, while the extension to the domain $z_0<0$ is straightforward. Due to symmetry, we can conclude that for any real $z_0\in (-\infty,+\infty)$ we have
\begin{equation}
    \mathcal{S}_u(z_0) = \erf \left( \dfrac{|z_0|}{\sqrt{2 \int_{0}^{u} \mathsf{T}_r e^{2\psi_r}dr}} \right).
\end{equation}
 \subsection{Survival probability in dimensionless time averaged over the initial distribution} The survival probability we have obtained depends on the starting point which is itself a random variable. By assuming that the starting position is sampled from a Gaussian  distribution, we have that the survival probability averaged over all  the initial positions reads
\begin{align}
    \mathcal{S}_u &=
     \int_{-\infty}^{+\infty} dz_0 \dfrac{\exp \left( -z_0^2/2\Sigma_0^2 \right)}{\sqrt{2\pi\Sigma_0^2}} \erf \left( \dfrac{\vert z_0\vert}{\sqrt{2 \int_{0}^{u} \mathsf{T}_r e^{2\psi_r}dr}} \right)\\
    &= \dfrac{2}{\pi} \tan^{-1} \left( \dfrac{\Sigma_0}{\sqrt{ \int_{0}^{u} \mathsf{T}_r e^{2\psi_r}dr}} \right),\label{eq:69}
\end{align}
where we recall that 
\begin{equation}
    \Sigma_0=\text{std}[Z_0]=\frac{\sigma_0}{\sqrt{\dfrac{2k_BT_{\rm c}}{\kappa_0}}},
\end{equation}
is the standard deviation of the initial position $\sigma_0=\text{std}[X_0]$ normalized by the characteristic length  $\sqrt{2k_BT_{\rm c}/\kappa_0}$, see Eq.~\eqref{eq:od_scaling}.  Note that, if the cycle is initialized from the equilibrium Boltzmann distribution, $\sigma_0=\sqrt{k_B T_{\rm c}/\kappa_0}$ and one has $\Sigma_0=1/\sqrt{2}$.
Considering that both the stiffness and the temperature are strictly positive functions, it is clear that the survival probability always decays to zero in the long time limit. Notably, the result~\eqref{eq:69} holds for any choice for the protocol for stiffness and temperature. In the next paragraph, we discuss further how $\mathcal{S}_u$ decays to zero for $u$ large  for the specific protocol used in the~GCE.
\subsection{Survival probability in physical time units}
Let us first express our formula for the survival probability~\eqref{eq:69} in the dimensional physical quantities:
\begin{equation}
    \mathcal{S}_t =
    \dfrac{2}{\pi} \tan^{-1} \left( \dfrac{\sigma_0/\sqrt{2k_BT_{\rm c}/\kappa_0}}{\sqrt{\int_{0}^{\frac{t}{\gamma/\kappa_0}} \mathsf{T}_r e^{2\psi_r} dr}} \right).
\end{equation}
Considering that the protocol, under the assumption that the particle has not crossed the origin in $[0,\tau/4]$, is given by the function
\begin{equation}
    \mathsf{K}_u = 1 + 4(\mathsf{K}_{\tau/2} -1) \left( \dfrac{\gamma/\kappa_0}{\tau} \right)^2 u^2
\end{equation}
thus, the function $\psi_{u} = \int_{0}^{u} \mathsf{K}_r dr$, see Eq.~\eqref{eq:psi}, reads
\begin{equation}
    \psi_{u} = u + 4(\mathsf{K}_{\tau/2} -1) \left( \dfrac{\gamma/\kappa_0}{\tau} \right)^2 \dfrac{u^3}{3},
\end{equation}
moreover in this first part of the protocol $T_u=T_{\rm c}$ thus $\mathsf{T}_u=1$.
Substituting this result in the formula for the survival probability we get, for any $0\leq t\leq\tau/4$,
\begin{equation}
    \mathcal{S}_t = 
    \dfrac{2}{\pi} \tan^{-1} \left[\frac{\sigma_0}{\sqrt{\frac{2k_BT_{\rm c}}{\kappa_0}}} \left( \frac{\kappa_0}{\gamma} \int_{0}^{t} \exp \left[2 \left( \frac{\kappa_0}{\gamma} t' + 4 (\kappa_{\tau/2}-\kappa_0) \frac{1}{\gamma \tau^2} \frac{t'^3}{3} \right)\right] dt' \right)^{-1/2} \right].
\end{equation}
If in the previous formula we denote $f_0 \equiv \kappa_0/\gamma$, $f_{\tau/2} \equiv \kappa_{\tau/2}/\gamma$ and we evaluate at $\tau/4$ we obtain
\begin{equation}
    \mathcal{S}_{\tau/4} =
    \dfrac{2}{\pi} \tan^{-1} \left[ \frac{\sigma_0}{\sqrt{\frac{2k_BT_{\rm c}}{\kappa_0}}} \left( f_0 \int_{0}^{\tau/4} \exp\left[2 \left( f_0 t + 4 \frac{f_{\tau/2} - f_0}{3 \tau^2} t^3\right) \right] dt \right)^{-1/2} \right],\label{eq:survgen}
\end{equation}
which is our main result. 
For the specific choice $\sigma_0=\sqrt{k_{B}T_{\rm c}/\kappa_0}$, i.e. when the initial state is at thermal equilibrium, we get the simplified form of Eq.~\eqref{eq:survgen} which is given by
\begin{equation}
    \mathcal{S}_{\tau/4} =\dfrac{2}{\pi} \cot^{-1} \left[  \left(2 f_0 \int_{0}^{\tau/4} \exp\left[2 \left( f_0 t + 4 \frac{f_{\tau/2} - f_0}{3 \tau^2} t^3\right) \right] dt \right)^{1/2} \right].
    \label{eq:survgen2}
\end{equation}
Note that, since $\cot^{-1}(x)\leq \pi/2$, we have $\mathcal{S}_{\tau/4}\leq 1$, as expected.
For large values of the cycle time $\tau$, the survival probability roughly decays to zero quicker than an exponential.


\section{Average work scaling: Gambling step}
\label{sec:E}
Formally, we can write $\langle W_{\rm I}\rangle_{\rm G}$ as ---we recall readers the definition of the gambling time, $\taug=\min(\mathcal{T}_0,\tau/4)$ with $\mathcal{T}_0 = \inf \{t\geq 0:X_t=0\}$ see Eq.~\eqref{stop_time} in the Main Text---
\begin{align}
    \label{work_def}
    \langle W_{\rm I} \rangle_{\rm G} 
    %
    &= \int_{-\infty}^{\infty} dx_0  P(x_0,0)\int_{0}^{\infty} d\mathcal{T}_0 \, \wp(\mathcal{T}_0|x_0) \int_{0}^{\min(\mathcal{T}_0,\tau/4)} dt\dfrac{1}{2} \dot{\kappa}_t \langle x^2_t\vert x_0,\mathcal{T}_0\rangle.  
\end{align}
Here, $P(x_0,0)=\exp\left(-x_0^2/(2\sigma_0^2)\right)/\sqrt{2\pi\sigma_0^2}$ is the distribution of the starting position and $\wp(\mathcal{T}_0|x_0)$ is the distribution of the first-passage time $\mathcal{T}_0$, and the other quantities in the integrands are described below. More precisely, the average~\eqref{work_def}  involves three quantities that may depend on the cycle duration $\tau$: 
\begin{enumerate}
\item The derivative of the stiffness with respect to time $\dot{\kappa}_t$,
\item The conditional first-passage-time probability density $\wp(\mathcal{T}_0|x_0)$, 
\item The average of the squared position  $\langle x^2_t\vert x_0,\mathcal{T}_0\rangle$ 
 evaluated at  times  $0\leq t\leq \min(\mathcal{T}_0,\tau/4)$ and conditioned to the initial position $x_0$ and to the first-passage time  to be $\mathcal{T}_0$.
\end{enumerate}
As we are going to show now, the two last terms tend to a constant non-zero value with respect to the cycle duration $\tau$ as it tends to $\infty$, whereas the first one scale as $1/\tau^2$. Eq.~\eqref{work_def} also involves an integration up to time $\tau/4$. Nevertheless, this integration does not affect the scaling, since $\wp(\mathcal{T}_0|x_0)$ decays exponentially with time. Thus when expanding for $\tau$ large we can safely extend the integral to $\infty$.
This overall explains the scaling found for $\langle W_{\rm I} \rangle_{\rm G}$.

\subsection{Time derivative of the stiffness} 
Along step I, we have $\kappa_t = \kappa_0 + \alpha t^2$ for all $0\leq t\leq \tau/4$, with $\alpha=4(\kappa_{\tau/2}-\kappa_{0})/\tau^2$.  Therefore,
\begin{equation}
\dot{\kappa}_t =  \frac{8(\kappa_{\tau/2}-\kappa_{0})}{\tau^2}t.
\label{eq:kappatsc}
\end{equation}
Because $\kappa_{\tau/2}-\kappa_{0}$ is fixed and thus independent of $\tau$, we have that $\dot{\kappa}_t$ scales like $\tau^{-2}$ with the cycle time.

\subsection{First-passage-time distribution} To access the first-passage time (FPT) distribution, we take the time derivative of the analytical expression for the survival probability [see Supplemental Material Sec.~\ref{sec:D} and Eq.~\eqref{eq:sprob}]
\begin{equation}
    \wp(\mathcal{T}_0|x_0, t_0) = -\left.\dfrac{\partial}{\partial t} \mathcal{S}_t(x_0, t_0)\right\vert_{t=\mathcal{T}_0} = \dfrac{\kappa_0}{\gamma} \dfrac{|z_0| \exp(2 \psi_{w,w_0})}{\sqrt{2 \pi \varphi_{w,w_0}^3}} \exp \left( -\dfrac{z_0^2}{2 \varphi_{w,w_0}} \right),
\end{equation}
where
\begin{equation}
    w = \frac{\mathcal{T}_0}{\gamma/\kappa_0},\quad w_0 = \frac{t_0}{\gamma/\kappa_0},
\end{equation}
are respectively the first-passage time and the initial time scaled by the relaxation time $\gamma/\kappa_0$. 
We also used here the dimensionless variable $z_0 = x_0/\sqrt{2k_BT_{\rm c}/\kappa_0}$, moreover, the functions $\psi$ and $\varphi$ read [cf. Sec.~\ref{sec:D}]
\begin{eqnarray}
    \psi_{w,w_0} &=& (w-w_0) + 4(\mathsf{K}_{\tau/2} -1)  \dfrac{(w-w_0)^3}{3}\left( \dfrac{\gamma/\kappa_0}{\tau} \right)^2,\label{eq:84}\\
     \varphi_{w,w_0} &=&  \int_{w_0}^{w} \exp \left[ 2(r-w_0) + 8(\mathsf{K}_{\tau/2} -1) \dfrac{(r-w_0)^3}{3}  \left( \dfrac{\gamma/\kappa_0}{\tau} \right)^2 \right]dr,\label{eq:85}
\end{eqnarray}
with $\mathsf{K}_{w}=\kappa_w/\kappa_0$.
In the large cycle time limit $\tau\gg \gamma/\kappa_0$, we have
\begin{eqnarray}
\psi_{w,w_0} &=& (w-w_0) \left[1+ A^{\psi}_{w,w_0} \left( \dfrac{\gamma/\kappa_0}{\tau} \right)^2\right] ,\\
     \varphi_{w,w_0} &=& \exp \left[ 2(w-w_0) \right] \left[1+B^{\varphi}_{w,w_0}\left(\dfrac{\gamma/\kappa_0}{\tau}\right)^2 
  +\,O\left( \left(\dfrac{\gamma/\kappa_0}{\tau}\right)^4 \right)\right]   .
\end{eqnarray}
Here, 
\begin{equation}
    A^{\psi}_{w,w_0}=(4/3)(\mathsf{K}_{\tau/2}-1)(w-w_0)^2,
\end{equation}
see Eq.~\eqref{eq:84}], and 
\begin{equation}
    B^{\varphi}_{w,w_0}= (1/3)(\mathsf{K}_{\tau/2}-1)\left(  \left( 2(w - w_0) \left( (w - w_0) \left( 2(w - w_0) - 3 \right) + 3 \right) - 3 \right) + 3 \right).
\end{equation}
In this limit, we find that the FPT distribution reads
\begin{equation}\label{eq:FPTsc} 
    \wp(\mathcal{T}_0|x_0, t_0) = \underbrace{\dfrac{\kappa_0}{\gamma} \dfrac{|z_0|  \exp \left[ 2(w-w_0) \right]}{\sqrt{\frac{1}{4} \pi ( \exp \left[ 2(w-w_0) \right] -1)^3}} \exp \left( -\dfrac{z_0^2}{e^{2 (w-w_0)} -1} \right)}_{\displaystyle \wp^{(0)}(\mathcal{T}_0|x_0, t_0)} + \,O\left( \left(\dfrac{\gamma/\kappa_0}{\tau}\right)^2 \right) 
\end{equation}
i.e., the first  correction to the FPT distribution  with respect to the quasistatic limit is of order $\tau^{-2}$.
Note that we use the superscript~$^{(0)}$ for all approximated quantities at zero-th order in $1/\tau$.
Note that the first order correction to the FPT distribution scales like $A/\tau^2$, but it is beyond our purpose to derive the explicit expression of  $A$.

\subsection{Conditional average of the squared position in an excursion} The last term whose scaling with $\tau$ we should analyze in the integrand of Eq.~\eqref{work_def} reads
\begin{equation}
    \langle x^2_t\vert x_0,\mathcal{T}_0\rangle = \int_{-\infty}^{+\infty} x^2 P(x,t| x_0,0; 0, \mathcal{T}_0) dx,
\end{equation}
where $P(x,t| x_0,0; 0, \mathcal{T}_0)$ is the probability density of being in $x$ at time $t$ given that the trajectory started in $x_0$ at time $0$ and crossed the origin for the first time at time $\mathcal{T}_0$. In other words, $P(x,t| x_0,0; 0, \mathcal{T}_0)$ represents the propagator of the so-called {\em excursion}~\cite{Majumdar_2015} associated with an Ornstein-Uhlenbeck process with time-dependent stiffness. Following Ref.~\cite{Majumdar_2015}, 
such a propagator is computed as follows
\begin{align}
    P(x,t|x_0, 0 ; 0,\mathcal{T}_0)&= \frac{P(x,t; 0, \mathcal{T}_0|x_0, 0)}{P(0,\mathcal{T}_0|x_0, 0)} \nonumber \\
	&=   \frac{P_a(x,t | x_0, 0)P(0,\mathcal{T}_0 | x, t)}{P(0,\mathcal{T}_0|x_0, 0)} \nonumber \\
	&=   \frac{P_a(x,t | x_0, 0)\wp(\mathcal{T}_0|x,t)}{\wp(\mathcal{T}_0|x_0,0)},
\end{align}
where in the equality step we split the joint probability density using the Markovianity of the process, and where $P_a$ is the propagator of the process with an absorbing boundary at the origin, which was introduced in Eq.~\eqref{eqn:P_a}. We report here the explicit expression for $P_a$
\begin{equation}
    P_a(x,t|x_0,0) = \dfrac{1}{\sqrt{\frac{2k_BT_{\rm c}}{\kappa_0}}}  \dfrac{\exp \left(\psi_{u,0}\right)}{\sqrt{2 \pi \varphi_{u,0}}} \left[ \exp\left(- \dfrac{(z e^{\psi_{u,0}}-z_0)^2}{2 \varphi_{u,0}} \right) - \exp\left(- \dfrac{(z e^{\psi_{u,0}}+z_0)^2}{2 \varphi_{u,0}} \right) \right],  
    \label{eq:99}
\end{equation}
where
\begin{equation}
    z = \frac{x}{\sqrt{\frac{2k_BT_{\rm c}}{\kappa_0}}},\quad  u=\frac{t}{\gamma/\kappa_0}.
\end{equation}
Equation~\eqref{eq:99} becomes, in the large cycle time limit,
\begin{equation}
   P_a(x,t|x_0,0) =\underbrace{\dfrac{1}{\sqrt{\frac{2k_BT_{\rm c}}{\kappa_0}}}  \dfrac{\exp(u)}{\sqrt{\pi (e^{2u}-1)}} \left[ \exp\left(- \dfrac{(z e^{u}-z_0)^2}{(e^{2u}-1)} \right) - \exp\left(- \dfrac{(z e^{u}+z_0)^2}{(e^{2u}-1)} \right) \right]}_{\displaystyle P^{(0)}_a(x,t|x_0,0)} + \,O\left( \left(\dfrac{\gamma/\kappa_0}{\tau}\right)^2 \right),
\end{equation}
in other words, the first order correction to $P_a$ scales, as the one for the FPT distribution, like $1/\tau^2$.
From this result, we find that the conditional average  of $x_t^2$ for trajectories that start in $x_0>0$ and cross the origin for the first time $\mathcal{T}_0$, evaluated at a previous time $t\leq\mathcal{T}_0$ obeys 
\begin{equation}
    \label{eq:x2scaling}
        \langle x^2_t\vert x_0,\mathcal{T}_0\rangle = \underbrace{\int_{0}^{\infty} dx \, x^2\, \frac{P_a^{(0)}(x,t | x_0, 0)\wp^{(0)}(\mathcal{T}_0\vert x,t)}{\wp^{(0)}(\mathcal{T}_0\vert x_0,0)}}_{\displaystyle \langle x^2_t\vert x_0,\mathcal{T}_0\rangle^{(0)}}+ \,O\left( \left(\dfrac{\gamma/\kappa_0}{\tau}\right)^2 \right),  
\end{equation}
i.e., its first order correction with respect to the quasistatic limit scales  also like  $\tau^{-2}$ with the cycle time.

\subsection{Scaling of the average work in the first stroke of the GCE}

Combining the scaling laws that we obtained [see Eqs.~\eqref{eq:kappatsc},~\eqref{eq:FPTsc}, and~\eqref{eq:x2scaling}]
\begin{equation}
    \dot{\kappa}_t =\,O\left( \left(\dfrac{\gamma/\kappa_0}{\tau}\right)^2 \right),\quad \wp(\mathcal{T}_0|x_0, t_0) = \wp^{(0)}(\mathcal{T}_0|x_0, t_0) +\,O\left( \left(\dfrac{\gamma/\kappa_0}{\tau}\right)^2 \right) ,\quad  \langle x^2_t\vert x_0,\mathcal{T}_0\rangle= \langle x^2_t\vert x_0,\mathcal{T}_0\rangle^{(0)} + \,O\left( \left(\dfrac{\gamma/\kappa_0}{\tau}\right)^2 \right),
\end{equation}
with the definition for $\expval{W_{\rm I}}_{\rm G}$ given by Eq.~\eqref{work_def}, we get that
\begin{equation}
    \langle W_{\rm I}\rangle_{\rm G} 
    =k_B T_{\rm c} \left(\frac{\tau_W}{\tau}\right)^2 + o\left(\tau^{-4} \right),
\end{equation}
where 
\begin{align}
\tau_W 
=&\sqrt{\frac{8(\kappa_{\tau/2}-\kappa_{0})}{k_B T_{\rm c}} \int_{-\infty}^{\infty} dx_0         P(x_0) 
 \int_{0}^{+\infty} d\mathcal{T}_0  \wp^{(0)}(\mathcal{T}_0|x_0, t_0)    \int_{0}^{\mathcal{T}_0} t \langle x^2_t\vert x_0,\mathcal{T}_0\rangle^{(0)} dt}
\end{align}
is independent of the cycle duration $\tau$. Notice that in the previous formula we could extend the integral over $\mathcal{T}_0$ to $\infty$ since, as specified before, $\wp(\mathcal{T}_0|x_0,t_0)$ decays exponentially fast with time.


\section{Additional figures}
\label{sec:F}

In this section, we display several figures containing additional numerical results supporting  the Main Text. 

\subsection{Heat exchanged with the cold reservoir} 
\begin{figure}[h!]
    \centering
\includegraphics[width=0.5\columnwidth]
{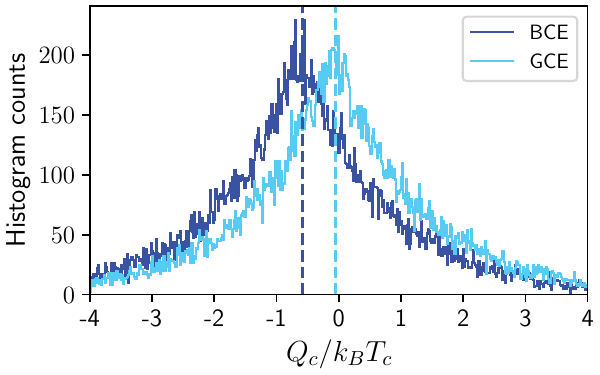}
    \caption{Histograms of the heat $Q_{\rm c}$ absorbed by the particle from the cold bath during the isothermal compression obtained from numerical simulations of the Gambling Carnot engine (GCE, light blue solid line), compared with the Brownian Carnot engine (BCE, dark blue solid line). Both histograms were obtained from $3\times10^4$ simulations of Eqs.~(\ref{eq:lang}-\ref{eq:TC}) with the gambling time defined as Eq.~\eqref{stop_time} for the GCE, while $\mathcal{T}=\tau/4$ for the BCE (no gambling). We set the cycle time to $\tau = 200$ms, and all other parameters same as in Fig.~\ref{fig2}. The vertical dashed lines are set to the empirical average $\langle Q_{\rm c} \rangle$ for each engine. 
     }
    \label{fig:heat-hist}
\end{figure}

Applying the first law of thermodynamics over a cycle to the definition of efficiency leads to equation~\eqref{eq:etaGW} in the Main Text, copied here for convenience,
\begin{equation*}
    \eta_{\rm G} 
    = 1 + \frac{\langle Q_{\rm c}\rangle_{\rm G}}{\langle Q_{\rm h}\rangle_{\rm G}}.
\end{equation*}
This result suggests that, $\eta_{\rm G}$ converges to one in the quasistatic limit whenever  the average heat exchanged with the cold reservoir $\langle Q_{\rm c}\rangle_{\rm G}$  converges to zero for large $\tau$. We show in Fig.~\ref{fig:heat-hist} empirical estimates of the distribution of $Q_{\rm c}$ (step I) for the GCE (cyan line in Fig.~\ref{fig:heat-hist}) obtained from $10^4$ numerical simulations with $100$kHz sampling frequency.  We also run simulations with the same statistics and parameter values for the BCE as comparison (blue line in Fig.~\ref{fig:heat-hist}), for which the mean of the distribution $\langle Q_{\rm c}\rangle<0$ as expected--heat is dissipated on average to the cold bath. Notably, our simulations reveal that $\langle Q_{\rm c}\rangle_{\rm G}=0$ within statistical errors, thus confirming that the GCE protocol renders the cold isothermal compression as an effective adiabatic process. In other words, the cold bath acts as an information reservoir during the isothermal compression of the GCE.

\subsection{Finite sampling effects} 
\begin{figure}[h!]
    \centering
    \includegraphics[width=0.54\linewidth]{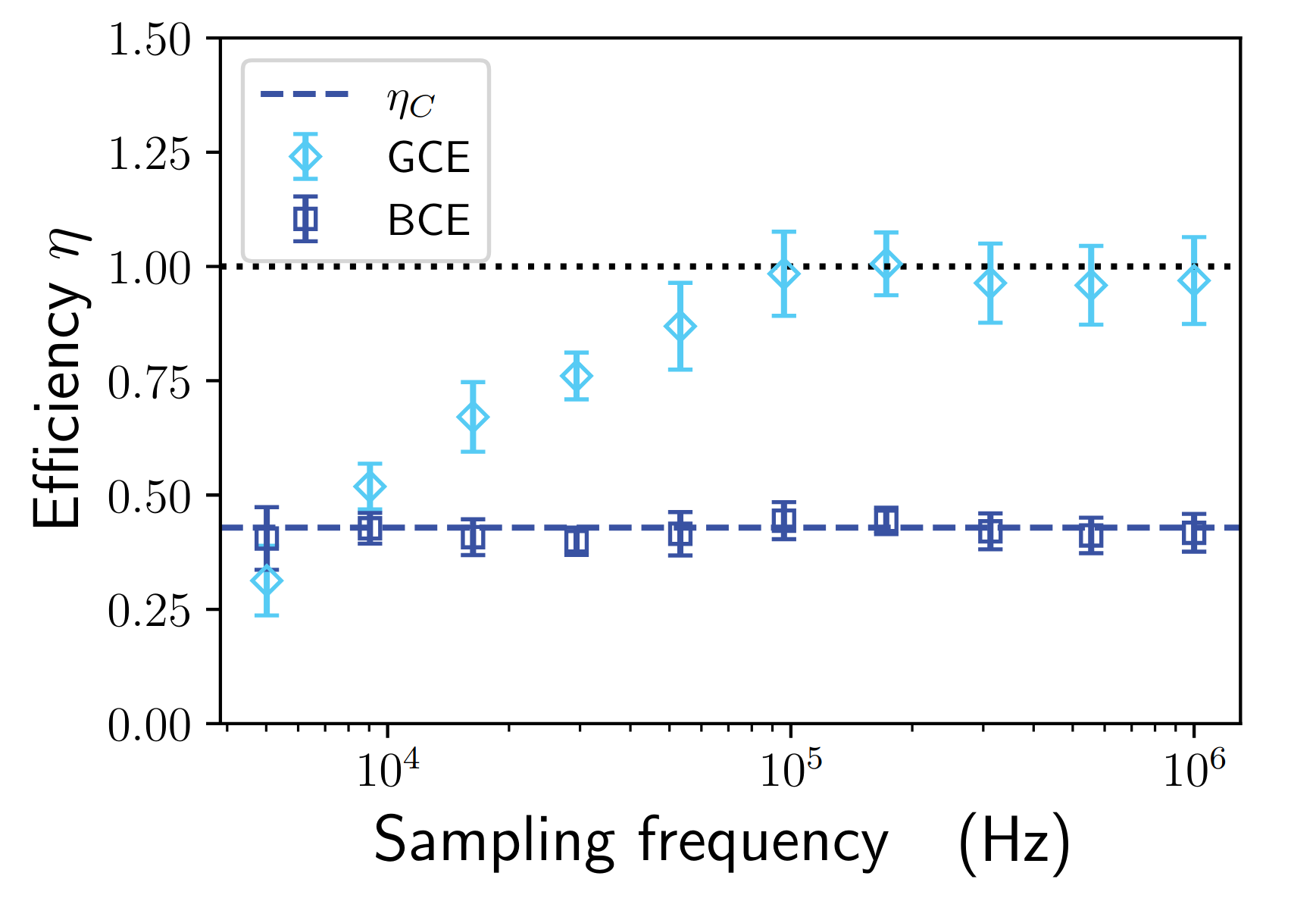}
    \caption{Efficiency of the GCE (cyan diamonds) compared with the BCE (blue squares) as a function of the  sampling frequency. Herein, unlike in the Main Text, we  include the work contribution associated with  the finite-time quench, see Eq.~\eqref{eq:workquench}. All parameters are set as in Fig.~\ref{fig2}, with $\tau=100$ms fixed for all the simulations. We used frequencies evenly spaced on a logarithmic scale between $5$KHz and $1$MHz. 
    }
    \label{fig:efficiency-sampling}
\end{figure}

Two key assumptions in the GCE's theoretical analysis are that the measurement of the particle's position is made with infinite precision, and that feedback takes place infinitely fast once the particle crosses the origin. A natural question  to ask is thus how the GCE's performance changes when these assumptions are relaxed. Figure~\ref{fig:efficiency-sampling} shows the dependence of the efficiency $-\langle W\rangle_{\rm G}/\expval{Q_{\rm h}}_{\rm G}$ at large $\tau$ for the GCE as a function of the  sampling frequency compared with that of the BCE $-\langle W\rangle/\expval{Q_{\rm h}}$. For this purpose we explored a wide range of sampling frequencies ranging from few kHz up to 1MHz, i.e. sweeping across all characteristic frequencies of the GCE.

Interestingly,  while the efficiency of the BCE is insensitive of the sampling frequency, the GCE's efficiency displays a strong dependency on the sampling frequency. More precisely, we observe a progressive decrease in the GCE efficiency up to a 50$\%$ of its maximum value when decreasing the sampling frequency  below $0.1$MHz up to few kHz. Above $0.1$MHz however, the data acquired is fine-grained enough to result in a value of the efficiency not significantly different from its saturating value at infinite bandwidth. 

The observed decrease in efficiency at lower sampling rates may be  explained intuitively as follows. By the time the particle is observed to have crossed the origin, the particle had sufficient time to move away from the origin by a  stochastic amount $X_{\mathcal{T}+(1/f)}$, with $f$ the sampling frequency. When a quench occurs, a nonzero amount of work is therefore exerted on the particle, which can be written as
\begin{equation}
    \langle \delta W_{\mathcal{T}} \rangle_{\rm G} =  \frac{1}{2}\left(\kappa_{\tau/4}\langle X_{\mathcal{T}+(1/f)}^2 \rangle_{\rm G} - \langle \kappa_{\mathcal{T}}X_{\mathcal{T}}^2 \rangle_{\rm G} \right).
    \label{eq:workquench}
\end{equation}
Because $\kappa_{\tau/4}\leq \kappa_{\taug}$ and $X_{\mathcal{T}+(1/f)}^2\geq X_{\mathcal{T}}^2$,  Eq.~\eqref{eq:workquench} implies that $\langle \delta W_{\mathcal{T}} \rangle_{\rm G}\geq 0$. In other words, subsampling leads to a work cost that results in a reduction of the overall efficiency, the cost being larger the smaller the sampling frequency.


\subsection{Range of validity of the overdamped approximation for the survival probability [Eq.~\texorpdfstring{\eqref{eq:sprob}}{TEXT} in the Main Text]} 

\begin{figure}[h!]
    \centering
    \includegraphics[width=0.46\linewidth]{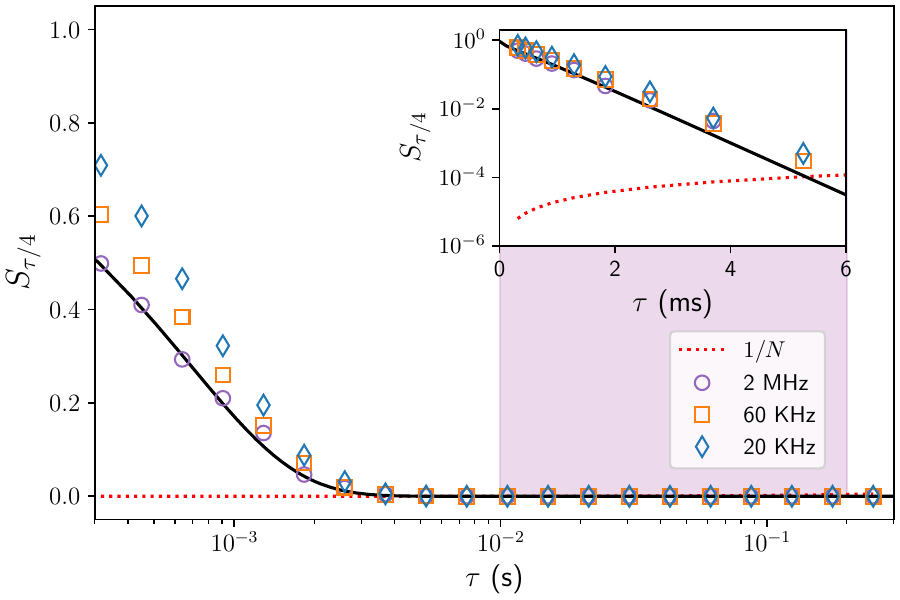}
     \includegraphics[width=0.46\linewidth]{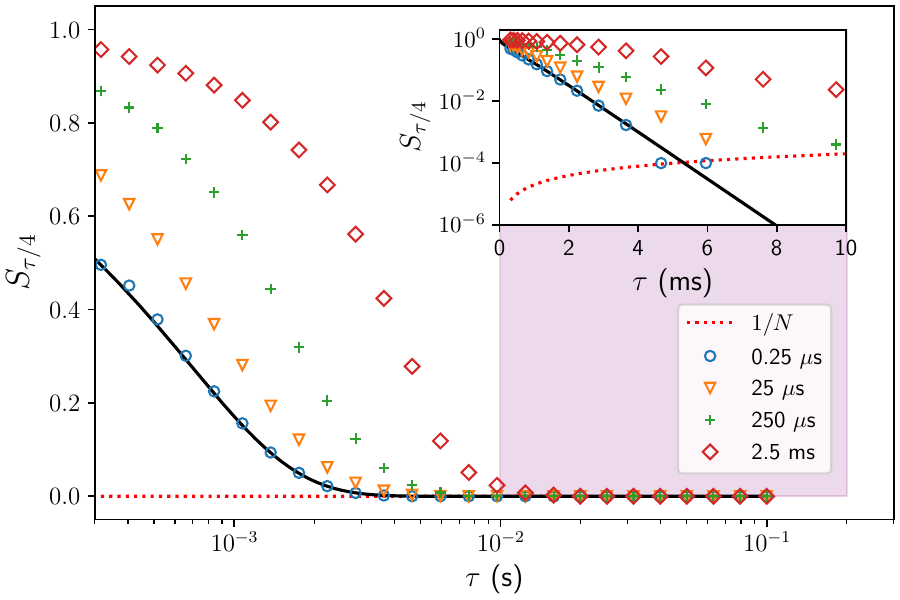}
    \caption{Left: Survival probability $S_{\tau/4}$ as a function of $\tau/4$, for an overdamped process \eqref{eq:langOD}. Points are simulated data, and the solid line is the analytical formula \eqref{eq:sprob}. Three different sampling frequencies are plotted (see legend), with the same parameter values as in figure \ref{fig2}, except for purple colored data, which was simulated with a sampling interval $\Delta t = 0.5 \mu$s, corresponding to a frequency of 2MHz. The red dotted line is $1/N_{\tau} = \tau / \tau_{\rm exp}$  a lower bound to the empirical survival probability detectable in a realistic experiment, assuming a single experiment lasts $\tau_{\rm exp}=50$s. Data were generated from $1.8\times 10^{4}$ trajectories. As for underdamped processes shown in the main text, figure \ref{fig:4}A, undersampling a trajectory leads to an overestimate of survival probability. Right: The survival probability $S_{\tau/4}$ \eqref{eq:sprob} as a function of $\tau/4$, for an underdamped process \eqref{eq:lang} at four different values of the momentum relaxation frequency $\tau_p$ (see legend), as defined in equation \eqref{eq:tau_p}. Underdamped simulation data (points) are in perfect agreement with the overdamped theoretical curve (black line)  at the value $\tau_p = 0.25 \mu$s, which corresponds to the experimental mass of the particle used in the original Brownian Carnot experiment \cite{martinez2016brownian}. The red dotted line is $1/N_{\tau} = \tau / \tau_{\rm exp}$ a lower bound to the empirical survival probability detectable in a realistic experiment, assuming a single experiment lasts $\tau_{\rm exp}=50$s. Parameter values as in figure \ref{fig2}, except for the simulation time step, which for this figure was $\Delta t = 0.2 \mu$s. Data were generated from $1.0\times 10^{4}$ trajectories.}
    \label{fig:sprob-add}
\end{figure}

Using low acquisition rates also results in an overestimation of the survival probability even if the dynamics is  overdamped in good approximation, see Fig.~\ref{fig:4}A in the Main Text.  We report in the left panel of Fig.~\ref{fig:sprob-add} estimates of the survival probability obtained from simulations of overdamped Langevin equations~\eqref{eq:langOD}. This would be equivalenet to setting $m=0$ in Eq.~\eqref{eq:lang} while keeping the rest of parameters the same as in the underdamped-dynamics simulations in Fig.~\ref{fig:4}A. 
We observe that our  analytical formula [Eq.~\eqref{eq:sprob} in the Main Text] fails to capture the effects of sub-sampling even if the underlying dynamics is overdamped. Such discrepancies are observable when using sampling frequencies of the order of $10$kHz and below, which are of the same order of magnitude as the characteristic corner frequencies $\kappa/\gamma$ in the overdamped Langevin equation~\eqref{eq:langOD}.   
We rationalize this effect by noting that our theoretical formula  [Eq.~\eqref{eq:sprob} in the Main Text] is derived under the assumption that time is continuous and we use infinite bandwidth sampling, therefore using low acquisition rates will necessarily delay the detection of a first-passage event. Such delay of detection explains the overestimation of the survival probability at low sampling rates, an effect that is again more pronounced the smaller is the sampling frequency.


The theoretical analysis  developed in Sec.~\ref{sec:D} for the survival probability \eqref{eq:sprob} assumes that for our parameter values and sampling frequencies, the dynamics of the particle and scaling behavior of $S_{\tau/4}$ may be well approximated by those associated with an overdamped Langevin equation~\eqref{eq:langOD}. The right panel in Fig.~\ref{fig:sprob-add} shows the impact of the momentum relaxation time $\tau_p=m/\gamma$ on the survival probability. We explored values of $\tau_p$ ranging from $0.25\mu$s (experimental value, used in the Main Text) up to $2.5$ms, by varying the particle mass $m$ while keeping the rest of the parameters the same as in previous simulations. Notably, when $\tau_p =0.25\mu$s, the overdamped approximation is reasonable because the momentum relaxation frequency $f_p=\gamma/m$ is of the order of MHz - three orders of magnitude larger than the position relaxation (corner) frequency of kHz. For heavier particles, e.g. $\tau_p=2.5$ms, we have $f_p\sim$kHz i.e. comparable with the position relaxation times. Within this range of larger masses $\tau_p\sim$ms and beyond, the overdamped approximation fails significantly and underestimates the actual survival probability, see  right panel in Fig.~\ref{fig:sprob-add}. Summarizing, our analytical expression for the survival probability is a reasonable approximation for light particles whose position is sampled quicker than the characteristic corner frequencies, as was the case in the experiment of the BCE~\cite{martinez2016brownian}.



\newpage


\begin{thebibliography}{46}
\bibitem [{not()}]{note}%
  \BibitemOpen
  \href@noop {} {}\bibinfo {note} {Here and further we will adopt the
  thermodynamic convention $W>0$ and $Q>0$ ($W<0$ and $Q<0$) when the work is
  exerted and heat absorbed (work is extracted and heat is dissipated) on the
  particle}\BibitemShut {NoStop}%
\end{thebibliography}
\end{document}